\documentclass[12pt,a4paper,authoryear,nofootinbib,preprint,titlepage]{elsarticle}
\usepackage{epstopdf}
\usepackage{bm}
\usepackage{latexsym}

\usepackage{rotating}

\usepackage{color}

\usepackage{amsmath,amsfonts,bm,amssymb}
\usepackage{amsthm}

\usepackage{eucal}
\usepackage{eufrak}

\usepackage{multirow}
\usepackage{rotating}

\usepackage{graphicx}
\graphicspath{{figures/}}
\usepackage{subfigure}

\newcommand{\captionfonts}{\footnotesize}

\makeatletter  
\long\def\@makecaption#1#2{%
  \vskip\abovecaptionskip
  \sbox\@tempboxa{{\captionfonts #1: #2}}%
  \ifdim \wd\@tempboxa >\hsize
    {\captionfonts #1: #2\par}
  \else
    \hbox to\hsize{\hfil\box\@tempboxa\hfil}%
  \fi
  \vskip\belowcaptionskip}
\makeatother   

\makeatletter
\renewcommand{\@seccntformat}[1]{{\csname the#1\endcsname}.\hspace{0.5em}}
\usepackage{setspace}

\begin{document}

\title{Why is equity order flow so persistent?}

\author[cfm,sfi]{Bence T\'oth\footnote{Corresponding author. E-mail: ecneb.htot@gmail.com. Tel.: +33 1 49 49 59 49}}
\author[monash]{Imon Palit\footnote{E-mail: imon.palit@monash.edu}}
\author[sfi,palermo,pisa]{Fabrizio Lillo\footnote{E-mail: fabrizio.lillo@sns.it}}
\author[oxford,sfi]{J. Doyne Farmer\footnote{E-mail: doynefarmer@gmail.com}}

\address[cfm]{Capital Fund Management, 23/25, rue de l'Universit\'e 75007 Paris, France}
\address[sfi]{Santa Fe Institute, 1399 Hyde Park Rd., Santa Fe NM 87501, USA}
\address[monash]{Department of Banking and Finance, Monash Univeristy, Melbourne, Australia}
\address[palermo]{Dipartimento di Fisica, Universit\'a di Palermo, Palermo, Italy}
\address[pisa]{Scuola Normale Superiore di Pisa, Piazza dei Cavalieri 7, 56126 Pisa, Italy}
\address[oxford]{Institute for New Economic Thinking at the Oxford Martin School and Mathematical Institute, 24-29 St. Giles, Oxford OX1 3LB}

\begin{abstract}
Order flow in equity markets is remarkably persistent in the sense that order signs (to buy or sell) are positively autocorrelated out to time lags of tens of thousands of orders, corresponding to many days.  Two possible explanations are herding, corresponding to positive correlation in the behavior of different investors, or order splitting, corresponding to positive autocorrelation in the behavior of single investors.  We investigate this using order flow data from the London Stock Exchange for which we have membership identifiers.  By formulating models for herding and order splitting, as well as models for brokerage choice, we are able to overcome the distortion introduced by brokerage.  On timescales of less than a few hours the persistence of order flow is overwhelmingly due to splitting rather than herding.  We also study the properties of brokerage order flow and show that it is remarkably consistent both cross-sectionally and longitudinally. 
\end{abstract}
\maketitle



\smallskip

 \smallskip


\section{Introduction}

\noindent Order flow in equity markets, defined as the process assuming value one for buyer initiated trades and minus one for seller initiated trades, is persistent in the sense that orders to buy tend to be followed by more orders to buy and orders to sell tend to be followed by more orders to sell.  Positive serial autocorrelation for the first autocorrelation of  order flow has been observed in many different markets\footnote{
Positive autocorrelation for a single lag was observed in the Paris Bourse by Biais, Hillion and Spatt (\citeyear{Biais95}), in foreign exchange markets by Danielsson and Payne (\citeyear{Danielsson12}), and in the NYSE by Ellul et al. (\citeyear{Ellul05}) and Yeo (\citeyear{Yeo06}).  See also \cite{Chordia02, Chordia05}.}.
In fact, order flow is {\it remarkably persistent}:  As illustrated in Figure~1, all the coefficients of the autocorrelation function of signed order flow are positive out to large lags, corresponding in trade time to tens of thousands of transactions or in real time to many days\footnote{
The extreme persistence of order flow was independently pointed out by Bouchaud et al. (\citeyear{Bouchaud04}) and \cite{Lillo03c}.  In fact order flow is so persistent that it is a long-memory process, i.e. its autocorrelation function is non-integrable (Beran, \citeyear{Beran94}).
This has been shown for the London and New York stock exchanges by \cite{Lillo03c}, for the Paris stock exchange by Bouchaud et al. (\citeyear{Bouchaud04}) and for the Spanish stock exchange by Vaglica et al. (\citeyear{Vaglica08}) and Moro et al. (\citeyear{Moro09}).
Note that \cite{axioglou11} argue that order flow is much more persistent within a given day than across successive days.  None of our results here depend on long-memory; we mention this only to emphasize the extreme persistence of order flow.
}.
 This is highly consistent across different markets, stocks, and time periods.

\begin{figure}[h!]
\begin{center}
\includegraphics[width=0.65\textwidth,angle=0]{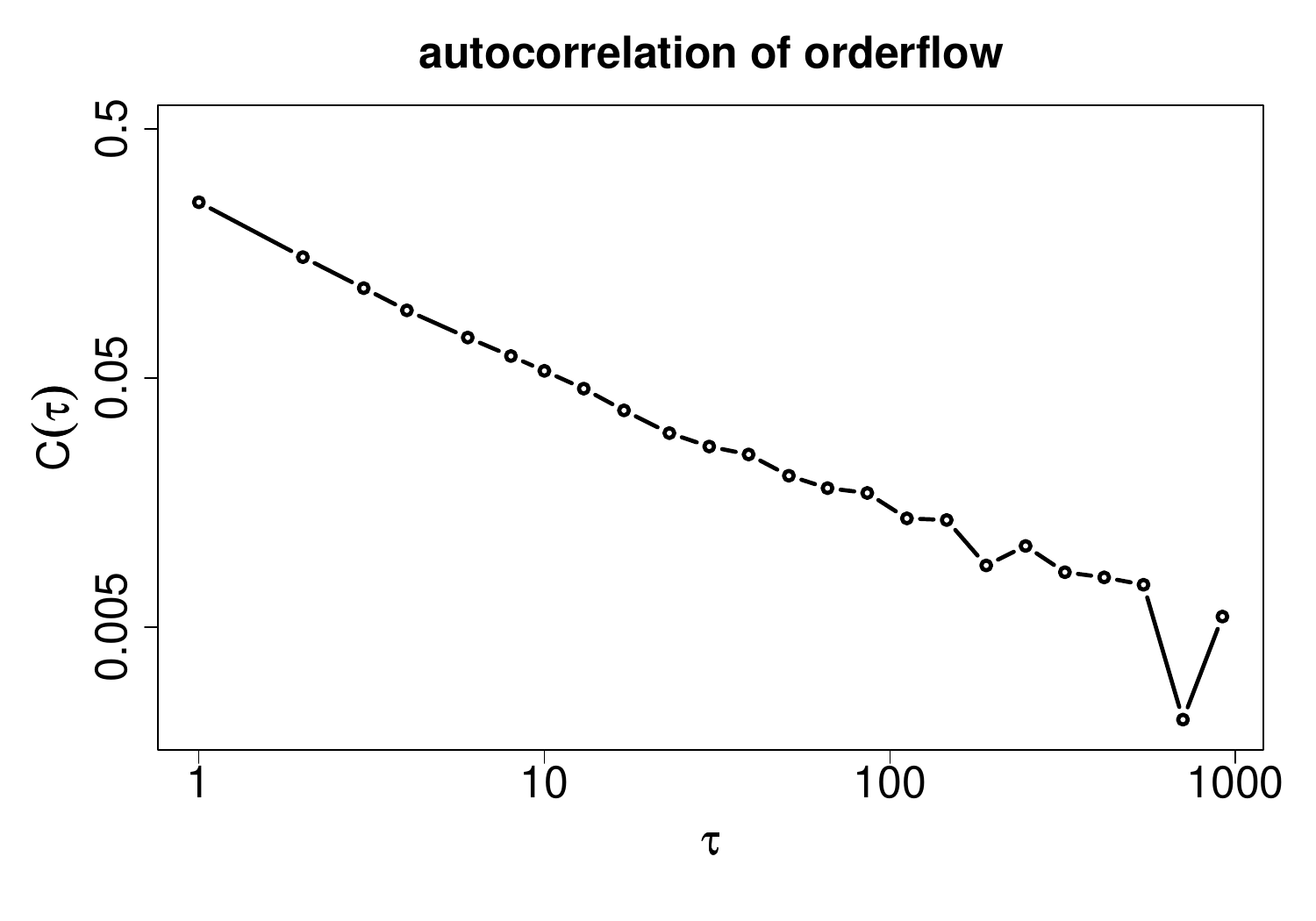}
\end{center}
\caption{\label{longmemFirst}  Autocorrelation function of order flow for the stock AZN in the first half of 2009, plotted on double logarithmic scale.  The time lag $\tau$ is measured in terms of number of effective market order placements, where an effective market order is any order that results in an immediate transaction.  To reduce estimation error we use order signs rather than order size and individual autocorrelations are binned for large lags.  The results are similar if we use order volume.}
\end{figure}

In this paper we perform an empirical study to elucidate the cause of this remarkable persistence.  This study is based on a unique data set from the London Stock Exchange (LSE) with codes indicating the exchange member who executed each order.   Members of the exchange may trade for their own accounts, but they may also act as brokers for investors who are not members of the exchange\footnote{
We often use the terms ``member" and ``broker" interchangeably.}.
As we will argue here, this provides useful information about the patterns of behavior of investors\footnote{
By {\it investor} we mean any trading entity with a coherent trading strategy.   This could correspond to a specific trading account within an institution such as an investment bank or hedge fund, or a private individual trading for his or her own account.},
even if it falls short of the fine grained data on the identity of investors that would make the results unequivocal.  

Our goal here is to distinguish between two fundamentally different types of behavior, order splitting and herding.  Order splitting occurs when single investors split desired large trades into smaller pieces and execute them gradually.  Our results here add to earlier evidence that order splitting is an important effect.  The strategic motivations for order splitting were originally derived by \cite{Kyle85}, who showed that an informed trader with a monopoly on private information would trade gradually in order to reduce impact.  Early empirical studies by \cite{Chan93,Chan95} using brokerage data with information about investors showed that order splitting is widespread.
\cite{Chordia02, Chordia05} found that daily order imbalances are serially autocorrelated and highly persistent\footnote{
Order imbalance for each stock over any time interval is typically calculated using the difference in the number of buy market orders and sell market orders, or the difference in the dollar value from buy market orders and sell market orders. Variations of this metric can use either a scaling factor or calculating a ratio instead of a difference. \cite{Brown97}, \cite{Chordia02}, \cite{Chordia04}, \cite{Chordia05}, \cite{Lo06}, \cite{Boehmer08} found correlation between order imbalance and returns, however, the evidences are not consistent on the direction of causality.  Looking at 5 minute order imbalances, \cite{Chordia08} have found that return predictability has declined over time.}, and pointed out that they can be caused either by order splitting or herding.
 Lillo, Mike and Farmer (\citeyear{Lillo05b}) introduced a model for order splitting connecting the size of large orders with the autocorrelation of order flow and showed that its predictions gave good agreement with data from the London Stock Exchange.  Gerig (\citeyear{Gerig07}), demonstrated that for the LSE stock AZN the trades coming from the same brokerage have long-memory, whereas trades from different brokerages do not (see also Bouchaud, Farmer, and Lillo (\citeyear{Bouchaud08b})).  Vaglica et al. (\citeyear{Vaglica08}) reconstructed the size of large orders from brokerage data and found that it is distributed as predicted by Lillo, Mike and Farmer.

An alternate hypothesis is that the extreme persistence of order flow is due to herding, as proposed by LeBaron and Yamamoto (\citeyear{Lebaron07,Lebaron08}), who constructed an agent-based model of imitation that produces highly persistent order flow.  There is a large literature on herding and its existence in equity trading at longer timescales is well-documented\footnote{
See for example \citep{Scharfstein90,Lakonishok92,Banerjee92,Banerjee93,Bikhchandani92,Hirshleifer94,Orlean95,Raafat09,Barber09}. Several previous studies have focused on the effect of communication network structure on price fluctuations;  see \citep{Kirman02,Iori02,Cont00b,Tedeschi09}.  \cite{Lakonishok92} and \cite{Wermers99} find only weak evidence for herding.}.
There are many strategic reasons why agents might herd, including reputational considerations, delayed response to public information, slow diffusion of private information, or imitation based on inferring the private information of others.  

Thus there are good arguments that one should expect both order splitting and herding.  Here we study the London Stock Exchange to quantitatively estimate their relative importance at short timescales, and in particular the extent to which these two mechanisms influence the observed persistent autocorrelation of order flow.  The paper that is most comparable to ours is \cite{Lee04}, who studied daily autocorrelations in order flow imbalances in the Taiwan Stock Exchange and inferred the presence of order splitting vs. herding. Their data contained identifiers for investors at a fine-grained level.  They aggregated the data into three categories corresponding to individuals, domestic institutions, or foreign institutions, further subdividing each of these into large or small, to create six groups in all.  They then studied the autocorrelations of the order flow of each group, comparing the results with or without deleting the trades of agents in the same group.  The change in the autocorrelation was used to infer splitting vs. herding, with results they interpreted as supporting both splitting and herding to varying degrees for each group.

The fact that our data contains only brokerage level data poses a greater challenge for data analysis.  Whereas Lee et al. (\citeyear{Lee04}) could aggregate their data using categories they were given by the exchange, a given brokerage potentially aggregates data from many different kinds of investors, and so it is not obvious {\it a priori} that this provides any useful granular information about single investors.  What we demonstrate here, is that at least for the question at hand, there is in fact enough information to make useful inferences.

We make both methodological and empirical contributions to the literature in several different ways.  First, we extend the work of Lee et al. (\citeyear{Lee04}) by formalizing their approach to decomposing order flow.  Second, we develop idealized models for order splitting and herding of single investors and compute the corresponding decomposition of order flow for each of these models.  Third, we introduce the concept of a {\it brokerage map}, which models how brokerages aggregate the orders of single investors.  Fourth, we combine the second and third contributions by computing the effect of brokerage for our models of individual investors.  Fifth, we apply this to empirically study the data, using our combined results in the fourth step to compare the data to well-formulated null hypotheses.  This allows us to place bounds on the relative importance of herding vs. splitting by individuals in causing the persistent autocorrelation of order flow.  

From an empirical point of view we contribute to the literature by performing a systematic study of a large high frequency data set containing brokerage identifiers. We systematically analyze six stocks in the London Stock Exchange over a period of ten years, a sample of more than 39 million orders.  We perform this analysis at the level of individual orders, on timescales ranging from that of individual order submissions up to a few hours (at longer timescales we begin to lose statistical significance in our results).  

Our main finding is that the dominant cause of persistence in order flow is the autocorrelated behavior of individual brokerages, rather than correlated order flow between brokerages. While there is some evidence for flows between brokerages, these are small compared with the persistence of order flow for individual brokerages.  By using the null hypotheses that we develop for investor behavior and brokerage we argue that this very likely reflects properties of single investors, and supports the hypothesis that order splitting is more important than herding at high frequencies\footnote{
Since obtaining the results presented here one of us (FL) has verified the main conclusions of the paper using a unique dataset for the Russian stock market which contains agent-level identifiers.}.

We wish to stress, however, that we are not disputing the well-documented existence of herding at low frequencies.  For example, \cite{Lu12} have used quarterly holdings of mutual funds to document herding in the LSE. This timescale is almost two orders of magnitude slower than that studied here, and was not focused on explaining the persistence of order flow. In addition our results do not contradict the earlier study\footnote{
\cite{Taylor03} did regressions for time series of count data, such as the number of transactions in a given interval, and found that lagged values had positive and statistically significant contributions.    Although he interpreted this as ``herding" he does not distinguish between herding and splitting as we do here, and since his data did not involve signed order flow, it relates more directly to trading volume than order imbalances.}
based on LSE data by \cite{Taylor03}.

Our results here have other interesting implications.  When we compare the data for different exchange members we find that for the largest members the behavior is surprisingly homogenous:  thirteen out of the fifteen largest members all have more or less identical correlations in their order flow.  Another interesting conclusion is that we are able to place bounds on the behavior of investors in choosing brokers.  In particular our empirical analysis, combined with our models of brokerage, makes it clear that investors work with only a small subset of brokers, and that dynamical variation in the choice of brokers is limited.  Finally we observe that the dominance of flows from the same brokers is constant across a nearly ten year period, despite the rise of algorithmic brokerage services during that period.

The cause of the persistence of order flow has important economic consequences.  Under the herding hypothesis the persistence of order flow is due to slow propagation of information, either slow reaction to public information or slow diffusion of private information, so that information is only gradually incorporated into prices.  In contrast, order splitting suggests that individuals receive private information (which in order to be new must be uncorrelated with past information arrival) and only trade gradually for strategic reasons.  As with herding, information is only slowly incorporated into prices, but for very different reasons.  Order splitting in the strong form that we observe here implies that the market deviates substantially from the equilibrium that would prevail if all participants were forced to reveal their true intentions at the outset of trading.

The persistence of order flow also has important economic consequences because it places strong constraints on the interaction of order flow and prices, which has implications for market impact\footnote{
For a comprehensive summary see \cite{Bouchaud08b}.}.
While the individual autocorrelation coefficients of order flow at each lag are small, the fact that they are all positive implies substantial predictability over long time horizons\footnote{
For an ARMA process predictability decays exponentially in time; for a long-memory process predictability decays as a power law, which is asymptotically much slower.}.  At the same time, in an efficient market the autocorrelation of price returns must be small.
\cite{Lillo03c} have argued that under the assumption that market impact is permanent this implies that liquidity is asymmetric between buying and selling in a time varying manner; Bouchaud et al. (\citeyear{Bouchaud04,Bouchaud04b}) have argued that this can also be resolved if impact is completely temporary, albeit decaying extremely slowly\footnote{
See also Farmer et al. (\citeyear{Farmer06}), \cite{Gerig07} and \cite{Bouchaud08b} for a detailed discussion of the sense in which these are equivalent.  More recent work by \cite{Eisler11} shows that when limit orders and cancellations are taken into account these models are not equivalent.}.
%

The presence or absence of order splitting affects the functional form of market impact, i.e. the dependence of price changes on order size.  \cite{Huberman04b} have shown that if liquidity is constant then permanent impact must be linear to avoid arbitrage; however, when liquidity is time varying, other functional forms become possible, see \cite{Gatheral10}.  The results cited above show that liquidity is indeed time varying, and empirical studies suggest that market impact is concave.  Under the assumption that trades are bundled by brokers and executed by order splitting algorithms, Farmer et al. (\citeyear{Farmer11}) show that the autocorrelation function of order flow determines the functional form of the market impact of large institutional orders, including both its permanent and temporary components.  Their derivation does not apply if the persistence of order flow is due to herding.  Thus our results here provide important background information relating to the nature of market impact.

The rest of this paper is organized as follows:   Section~\ref{data} describes the data.  In Section~\ref{autocorrelationSection} we derive a decomposition of the autocorrelation function of order flow into that due to the same agent vs. that due to other agents.   In Section~\ref{individualInvestors} we develop models for herding and splitting by investors, and compute their decomposition.  In Section~\ref{brokerageSection} we develop models for brokerage and compute how brokerage distorts the decomposition of autocorrelation when combined with the investor models developed in the previous section.  In Section~\ref{empiricalResults} we apply our methods to the data and demonstrate the dominance of the contribution to the autocorrelation of brokerages with themselves vs. other brokerages. We also test the data against the null hypotheses developed earlier and argue that this very likely reflects the behavior of investors.   We also study the heterogeneity of exchange members and show that most of them are extremely persistent in terms of both trade direction and the clustering in time of their trades.   In Section~\ref{antiHerding} we investigate the negative contribution of herding to the autocorrelation function more carefully, and show that it is driven by heterogeneity in the response of investors to market orders that change the price, versus those that do not. In the conclusions we summarize and reflect on the economic implications of our results.

\section{Data\label{data}}

This study is performed using data from the London Stock Exchange (LSE).   There are two parallel markets,  the on-book market (SETS) and the off-book market (SEAQ). In the on-book market trades take place via a fully automatic electronic order matching system, while in the off-book market trades are arranged bilaterally via phone calls. We restrict our study to the on-book market\footnote{Given also the long time span and the equity markets restructuring in recent years, it is difficult to estimate the fraction of volume traded on the on-book market.  Carollo et al. (\citeyear{Carollo12}) estimate that it ranges between 40\% and 57\% for nine highly liquid stocks traded at LSE in 2004.}.  Note that there are no official market makers, though it is possible for any member firm to act as a market maker by posting bids and offers simultaneously.

We study six stocks in the period from June 2000 to June 2009, with the exception of a six month period from January to May in 2003.   We divide the data for each stock into 17 subperiods of six months each, for a total of $17\times 6=102$ samples.  The six stocks we study are AstraZeneca (AZN), BHP Billiton (BLT), British Sky Broadcasting Group (BSY), Lloyds Banking Group (LLOY), Prudential (PRU), Vodafone Group (VOD)\footnote{
We have also investigated other stocks and found similar results.}.
In cases where we present data for only one period we will use AZN in the first half of 2009. In this period the typical number of market orders per day is between $4,000$ and $5,000$.   The number of market orders in each sample ranges from $91,710$  to $1,416,000$, with an average of $382,500$.

Our data set contains all orders placed in the on-book market.  The aspect of this data that makes it unique is that each order is characterized by a code identifying the member who placed the order\footnote{
We do not have the actual names of the members, but rather anonymized codes uniquely identifying each member.  For the period 2000-2002 we are only able to track members for one month, because in this part of the data codes are reshuffled each month. Because we are operating on timescales of at most a few hours this does not affect the results.}.
The number of members varies throughout the sample, but there are typically about 100 members.   To avoid data with poor statistics, in each six month period we remove members who make less than $100$ transactions across the full period, which typically leaves about $80$ active members.  The activity level is very heterogeneous.  For example, the 5 most active market members are responsible for 40-50\% of transactions and the 15 most active ones are responsible for 80-90\% of transactions.  The value of the Gini coefficient of member activity averaged across the 102 samples is $0.87$.  This is quite consistent across stocks and time periods, with a standard deviation of only $0.02$.   Thus the trading activity is strongly concentrated in a relatively small number of member firms.

We have performed the analysis given here in three different ways: (1) market order signs $\epsilon_t$, where a market order is defined as any event that results in an immediate transaction; (2) signs of all orders, including both market and limit orders; (3) signed volume $v_t$ of transactions.  The results are similar in all three cases, except that for signed volume they are somewhat noisier.  We present only the results for case (1), market order signs.  The number of market orders is used to measure time, i.e. $t'= t+1$ whenever there is a market order placement. As a final methodological note, we do not need to infer whether the trade is buyer or seller initiated by comparing the trade and the prevailing quote price, because our data contain explicitly the information on the sign of the trade. 

\section{Decomposing order flow\label{autocorrelationSection}}

In this section we present two decompositions of the autocorrelation of order flow that will serve as the basis for our empirical analysis.

Consider a time series of orders of sign $\epsilon_t$, where $\epsilon_t = + 1$ for buy orders and $\epsilon_t = - 1$ for sell orders.  For convenience we measure time in terms of order arrivals, so that time advances by one unit every time a new order arrives.  We define the signed order flow $\epsilon_t^i$ to be zero if the order at time $t$ was not placed by agent $i$ and to be the order sign otherwise, i.e.
\begin{eqnarray*}
\epsilon_t^i & = & ~~1~=> ~\mbox{buy order placed by agent}~i\\
\epsilon_t^i & = & ~~0~=> ~\mbox{order placed by another agent}\\
\epsilon_t^i & = & -1~=> ~\mbox{sell order placed by agent}~i.
\end{eqnarray*}
For now we leave it open whether the agent is an individual investor or whether the agent is a brokerage; this will become clear in context.  As already mentioned in Section \ref{data} the analysis can be performed using market orders (as we have done here), all orders, or with signed order volumes, with similar results.

Let $N$ be the length of the time series, $N^i$ be the number of orders due to agent $i$ and $N^{ij}(\tau)$ the number of times that an order from agent $i$ at time $t$ is followed by an order from agent $j$ at time $t + \tau$.
Similarly, let  $P^i=N^i/N$ be the fraction of orders placed by agent $i$, and $P^{ij}(\tau)=N^{ij}(\tau)/N$ be the fraction of times that an order from agent $i$ at time $t$ is followed by an order from agent $j$ at time $t + \tau$. The weighting function $P^{ij}(\tau)$ captures the extent to which agents $i$ and $j$ tend to be active with a given time lag $\tau$, regardless of the sign of their orders.  If agents act independently then in the large $N$ limit:
\begin{equation}
\nonumber
P^{ij}(\tau)=P^i P^j~~~~~\forall i,j,
\end{equation}
and $P^{ij}(\tau)$ is independent of $\tau$.
Deviations from independence can be characterized by
\begin{equation}
\tilde P^{ij}(\tau) \equiv P^{ij}(\tau)-P^iP^j.
\label{tildep}
\end{equation}
Finally, let the sample mean of the order sign for agent $i$ be
\[
\mu^i \equiv \frac{1}{N^i}\sum_t\epsilon_t^i.
\]
The sample autocorrelation in the order signs of agents $i$ and $j$ can then be written
\begin{equation}
\label{autocoreDefinition}
C^{ij}(\tau) \equiv  \frac{1}{N^{ij}(\tau)} \sum_{t}^{}  \epsilon_t^{i} \epsilon_{t + \tau}^j-\mu^i \mu^j.
\end{equation}

\subsection{Decomposition by agent identity}

As shown in the Appendix, the autocorrelation function of order flow can be written
\begin{equation}
C(\tau)=\sum_{i,j} P^{ij}(\tau)C^{ij}(\tau)+\sum_{i,j} \tilde P^{ij}(\tau) \mu^i \mu^j .
\label{ptilde}
\end{equation}
This can trivially be written in the form
\begin{equation}
C(\tau) =C_{same}(\tau)+C_{other}(\tau),
\label{autocdecomp}
\end{equation}
where
\begin{eqnarray}
\label{components}
&&C_{same}(\tau) \equiv \sum_{i} P^{ii}(\tau)C^{ii}(\tau)+\sum_{i} \tilde P^{ii}(\tau) (\mu^i)^2,\\
\nonumber
&&C_{other}(\tau) \equiv \sum_{i\ne j} P^{ij}(\tau)C^{ij}(\tau)+\sum_{i \ne j} \tilde P^{ij}(\tau) \mu^i \mu^j.
\nonumber
\end{eqnarray}
$C_{same}(\tau)$ includes only the diagonal contributions, and is the autocorrelation of orders coming from the same agent.  In contrast $C_{other}(\tau)$ includes the off-diagonal contributions and so is the autocorrelation of orders coming from different agents.   As we will argue in the next section this is related to the relative contributions of splitting vs herding.

Eq.\ (\ref{autocdecomp}) can be trivially rewritten as
 \begin{equation}
 \nonumber
 \frac{C_{same}(\tau)}{C(\tau)} + \frac{C_{other}(\tau)}{C(\tau)} = 1.
 \end{equation}
Thus all the information about the decomposition is contained in the ratio
 \begin{equation}
S(\tau)=\frac{C_{same}(\tau)}{C(\tau)} = 1 - \frac{C_{other}(\tau)}{C(\tau)}.
\label{s}
\end{equation}
For convenience we will sometimes characterize the decomposition of order flow based on agent identity in terms of $S(\tau)$.


\subsection{Decomposition by correlation of trading sign vs. timing of activity}

It is useful to compare the first and second terms of the decomposition in Eq.~(\ref{ptilde}).  The first term weights the correlation of order signs by the activity $P^{ij}(\tau)$, and the second term weights deviations in activity by the mean order signs $\mu^i$ and $\mu^j$.  Since buying and selling roughly balance, for large samples $\mu^i$ is small and the second term is negligible\footnote{
It is not surprising that this term is small --  in the long-run one expects $\mu^i$ to be close to zero.  If $\epsilon^i_t$ represents signed trading volume, for example, then $\sum_t \epsilon^i_t$ is the total inventory change during the period of the study.  If the inventory remains bounded, then $\mu^i \to 0$ in the limit $N \to \infty$.  Even if $\epsilon^i_t$ represents order signs, then if agents buy as often as they sell, $\mu^i$ is small.}.
Typical values for our dataset are $|\mu^i| = 0.03$ and  $\sum_{i,j} \tilde P^{ij}(1) \mu^i \mu^j = 10^{-4}$.  %
Thus the first term of Eq.~(\ref{ptilde}) is three orders of magnitude larger than the second one, and it is a very good approximation to take
\begin{equation}
C(\tau) \simeq \sum_{i,j} P^{ij}(\tau)C^{ij}(\tau).
\label{Capprox}
\end{equation}
Similar approximations can be used for $C_{same}(\tau)$ and $C_{other}(\tau)$.   One can think of the autocorrelation as a product of two terms, one that depends on activity, as reflected by $P^{ij}(\tau)$, and the other that depends on trading direction,  as reflected by $C^{ij}(\tau)$.

We have thus decomposed order flow in two different ways:  Eq.~(\ref{components}) decomposes the total autocorrelation $C(t)$ into the contribution $C_{same}$ due to the same agent  vs. the contribution $C_{other}$ due to different agents.  In contrast, Eq.~(\ref{Capprox}) decomposes contributions due to trading direction (buying versus selling) versus activity clustering (when agents make their trades).

\section{Models for splitting and herding\label{individualInvestors}}

In order to use the decompositions developed in the previous section to understand the relative importance of splitting vs. herding we first need to understand how these agent behaviors affect the decomposition of the autocorrelation into same vs. other.  At first glance this might seem to be trivial.  Splitting necessarily involves autocorrelations induced by the same agent, so if $C_{same} > C_{other}$ then one would naturally infer that splitting is the dominant cause of the persistence of order flow.  Similarly, since herding involves interactions of agents with each other, then if $C_{other} > C_{same}$ it would seem that we could naturally infer that herding is the dominant cause. However, things are not quite so simple. The basic reason is that herding can involve complicated feedback between agents, who generate a mixture of correlations with each other and with themselves.

To make this clearer we formulate explicit models of both splitting and herding.
The model of splitting is straightforward, but there are many ways to model herding, corresponding to the many forms in which herding can occur.  It is useful to distinguish between herding based on a common response to public information and herding based on the endogenous dynamics of investors who are responding to each others' private information.   We provide an example of each.   There are of course many variations; we make some arbitrary choices, aiming for simplicity.  As we will argue later by comparing several different models for herding, the results are very robust and the main conclusions are not sensitive to the details.

For the purposes of this section one should interpret an agent as a single investor.  As stated in the introduction, by single investor we mean an agent who is trading for a given account using a well defined and consistent strategy (in contrast to a broker -- see the next section).

\subsection{Order splitting}

Under order splitting the autocorrelation of order flow is generated entirely by investors who split large trades into small pieces and execute them incrementally.  Under the assumption that all persistence is due to order splitting, we assume that information is private and IID.  Since by assumption the autocorrelation between different investors is zero, the persistence is entirely due to splitting, and
\begin{eqnarray}
\nonumber
C_{same} (\tau) & = & C(\tau),\\
C_{other} (\tau) & = & 0.
\label{orderSplittingCor}
\end{eqnarray}
Furthermore order splitting implies $C_{same}(\tau) > 0$, as discussed in \cite{Lillo05b}.

If the information arrival is not IID it is of course possible that $C_{same}(\tau) > 0$ due to a cause other than splitting, such as positive autocorrelation in information arrival.  If we can only observe trades and have no further information about the information or intentions of the investors then we will not be able to make this distinction.  The reader should bear this in mind:  Order splitting is an example of a process that causes $C(\tau) = C_{same} (\tau)$, but it is not the only example.

\subsection{\label{herdingModel1} Herding model I:  Public information}

This model captures the idea of herding as a response to a common external cause, i.e. a herd that is directed from outside by public information.  It has the important advantage of being trivial to compute.  

Assume an information signal $I_{t'}$ taking on positive and negative integer values.  The sign tells if investors should buy or sell and the absolute value tells how many investors are affected by the new information. Specifically, the signal is received by $I_{t'}$ investors, who are chosen randomly (with replacement) from a pool of $M$ possible investors with probability $P^i > 0$, where $\sum_{i=1}^M P^i = 1$.  Each of the investors submits an order of size one whose sign is equal to the sign of $I_{t'}$.  This is repeated at each real time step $t'$, so that when measured in terms of order arrival time $t$ the autocorrelation of the order flow is $C(\tau)$.  To the extent that $I_{t'}$ is broadcast to multiple investors the information is public\footnote{
As stated, for blocks of length $I_{t'}$ this model generates sequential orders all of the same sign.  It is possible to make the order flow look more realistic by injecting orders with a random sign, but the only effect is to decrease the prefactor of the autocorrelation without otherwise affecting its time dependence.  The order flow autocorrelation $C(\tau)$ depends on the autocorrelation of $I_{t'}$ and its size distribution $P(I_{t'})$, but these do not affect the decomposition.}.

The decomposition of Eq.~(\ref{components}) is easily derived in closed form.  Since the investors are chosen randomly the timing and sign of the trade are independent of the investor, which implies
\begin{eqnarray}
\label{independence}
C^{ij}(\tau) & = & C(\tau),\\
\nonumber
P^{ij}(\tau) & = & P^iP^j.\
\end{eqnarray}
The second of these relations implies that  $\tilde{P}^{ij}(\tau) = 0$, $\forall i,j$.  Eq.~(\ref{components}) then implies that
\begin{eqnarray*}
\nonumber
C_{same} (\tau)= \sum_{i}P^{ii}(\tau)C^{ii}(\tau)=C(\tau)\sum_{i}(P^{i})^2\\
C_{other}(\tau) = \sum_{i\ne j}P^{ij}(\tau)C^{ij}(\tau)=C(\tau)\sum_{i\ne j}P^{i}P^j.
\end{eqnarray*}
Defining the {\it investor heterogeneity} $\mathcal{H}$ as the cross-sectional variance of the investor trading frequencies $P^i$,
\begin{eqnarray}
\label{heterogeneityDefinition}
\mathcal{H} = \mbox{Var}[P]=\frac{1}{M}\sum_{i}(P^{i})^2-\frac{1}{M^2},
\label{varP}
\end{eqnarray}
and substituting for $\mathcal{H}$ gives
\begin{eqnarray}
\label{eq:anal}
\nonumber
C_{same}(\tau) & = & C(\tau) \Big(\frac{1}{M}+M \mathcal{H} \Big),\\
C_{other}(\tau) & = & C(\tau) \Big(\frac{M-1}{M}-M \mathcal{H}\Big).
\end{eqnarray}

\begin{figure}[tb]
\begin{center}
\includegraphics[width=0.65\textwidth,angle=0]{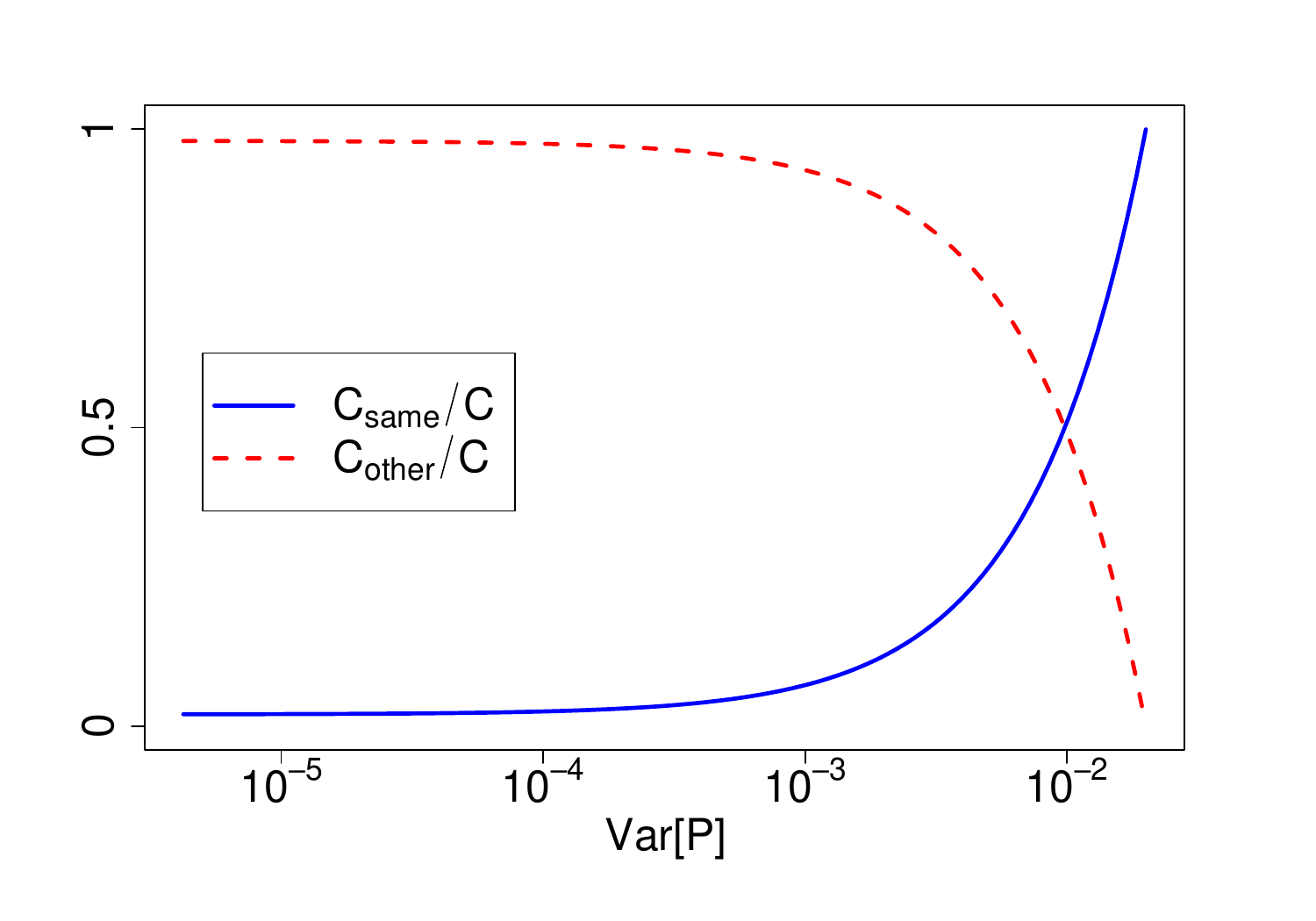}
\end{center}
\caption{\label{herding1}The relative contribution $C_{same}/C$ vs. $C_{other}/C$ for the simple herding model, plotted as a function of the investor heterogeneity $\mathcal{H} = \mbox{Var}[P]$.  $C_{other}/C$ might normally be interpreted as the ``herding contribution"; this is at its maximum when the investor heterogeneity is zero, and at its minimum when all the trading is concentrated on a single investor.  In contrast for order splitting $C_{same} = C(\tau)$ independent of the investor heterogeneity.}
\end{figure}

The investor heterogeneity $\mathcal{H}$ lies in the range $0 \le \mathcal{H} \le \left(1 - 1/M \right)/M$.   For a given $M$ the lower bound occurs when all investors are equally active, i.e. $\mathcal{H} = \mbox{Var}[P] = 0$, where
\begin{eqnarray*}
C_{same} & = & \frac{C(\tau)}{M},\\
C_{other} & = &  \frac{M-1}{M}C(\tau).\\
\end{eqnarray*}
Note that in this case, in the limit $M \to \infty$, $C_{same}(\tau) \to 0$ and $C_{other}(\tau) = C(\tau)$.  That is, providing the pool of investors is sufficiently large, the autocorrelation is driven entirely by $C_{other}(\tau)$.  The upper bound occurs when almost all trading is concentrated in a single investor\footnote{
Note that when trading is concentrated on a single investor the value of $M$ is no longer well defined, so we take the limit $P^1 \to 1$.  In this limit the heterogeneity is at a maximum in the sense that the difference between the trading activity of investors is as large as it can be.  We recover $C_{same} = C(\tau)$ as expected.},
i.e. $\mathcal{H} = \mbox{Var}[P] = \left(1 - 1/M \right)/M$, where
\begin{eqnarray*}
C_{same} & = & C(\tau),\\
C_{other} & = & 0.
\end{eqnarray*}
If $C(\tau) > 0$ then $C_{same}$ is always positive and $C_{other}$ is always non-negative.

As the trading goes from uniformly distributed to concentrated in a single member, the relative contribution $C_{same}/C$ grows from $0$ to $1$ while $C_{other}/C$ decreases from $1$ to $0$, as shown in Figure~\ref{herding1}.  Thus for this simple herding model there is generally a non-zero contribution to $C_{same}$, which grows with investor heterogeneity.  This is because there is a nonzero probability that the same agent will trade more than once during the time it takes for the correlation to decay.

\subsection{Herding model II:  Private information with imitation}

Our second herding model produces herding endogenously without need for an external coordination mechanism.  In this model the investors herd by imitating each other, and the autocorrelation arises from their collective behavior.   We assume that investors exist within a social network where information is transmitted between neighbors. The transmission of information depends on the topology of the network, which also determines the autocorrelation of order flow. 

\begin{figure}[tb]
\begin{center}
\includegraphics[width=0.65\textwidth,angle=0]{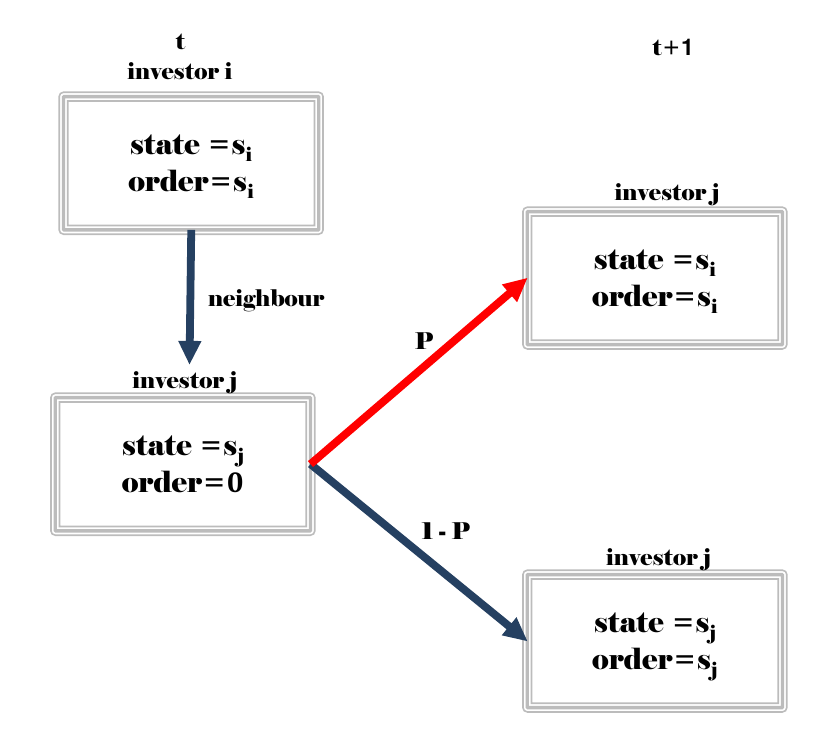}
\end{center}
\caption{\label{schematic}{\it Schematic diagram of herding model II.}  Assume a fully connected social network.  At time $t$ an agent is chosen randomly with probability $P^i$, who submits an order $s_i = \pm 1$, where $s_i$ is her state.  At time $t+1$ neighbor $j$ of agent $i$ is chosen at random.  With probability $p$ agent $j$ imitates agent $i$ by submitting order $s_i$ and changing state to $s_i$, and with probability $1 - p$ she ignores agent $i$, submitting order $s_j$ and keeping her original state $s_j$.  This continues for times $t + 2$, $t+3$, etc., choosing all the neighbors without replacement as needed.  At this point a new agent $i$ is chosen at random and the process is repeated. }
\end{figure}

The nodes of the social network correspond to investors and links correspond to influence.  For a schematic diagram of the model see Figure~\ref{schematic}.  
The network is connected, i.e. all nodes are joined to a common graph by at least one undirected link.  Investor $i$ has a binary state $s^i_t = \pm1$ that indicates whether, all else equal, this investor prefers to buy or sell. However, with probability  $p\in[0,1]$ she may choose to imitate her neighbor.

The order flow is generated using the following algorithm:
\begin{enumerate}
\item
At time $t$ an investor $i$ is chosen randomly and an order with sign $s^i_t$ is submitted.
\item
$t \to t+1$.  Consider a neighbor $j$ of investor $i$.  With probability $p$ she imitates $i$, and with probability $(1-p)$ she trades on her own information.  In more technical terms, with probability $p$ investor $j$ submits an order $\epsilon^j_t = s^i_t$ and changes state to $s^j_{t+1} = s^i_t$, and with probability $1-p$ she submits an order $s^j_t$ and preserves her state, i.e. $s^j_{t+1} = s^j_t$.
\item
If all neighbors of $i$ have been considered go to step (1); otherwise go to step (2) and choose one of the remaining neighbors (without replacement).
\end{enumerate}
The degree distribution $\mathcal{P}(\ell)$  and the parameter $p$ determine the autocorrelation $C(\tau)$ of the order flow\footnote{
The degree of a node of a graph is the number of links connected to that node}.

This model has two fixed points, corresponding to the case that all the agents are in the same state, either $s^i = +1$ or $s^i = -1$ for all $i$.   However, providing the system is large and is initially placed in a sufficiently diverse initial configuration, these fixed points cannot be reached in a reasonable time, and the system will generate time correlated but otherwise random order flow for a very long time.  It is also possible to generalize the model to inject external information by occasionally randomly altering the sign of note $i$ in step (1), in which case the system will never settle into a fixed point.  Providing the rate of injection of external information is low, this makes a negligible difference in $C(\tau)$.

For this model we are not able to compute the decomposition $(C_{same}, C_{other})$ in closed form. Instead we simulate it and compare it to the public information herding model, tuning the parameters of herding model II so that the autocorrelation function $C(\tau)$ is the same as that of model I.   To produce a highly persistent autocorrelation function we construct a social network using preferential attachment \citep{Simon55, Barabasi99}.   An initial node is created, and then new nodes are incrementally generated and connected to a randomly chosen pre-existing node with a probability of attachment  proportional to its degree.  Because we connect to only one pre-existing node at a time the resulting graph is a tree\footnote{
We could more realistically attach to multiple pre-existing nodes, but this does not affect the results.}.
For the private information herding model the persistence of order flow derives from the heavy tailed degree distribution of the social network.  The process of imitation converts the scale free power law degree distribution of the social network, which by construction is of the form $\mathcal{P}(\ell) \sim \ell^{-\eta + 1}$, into order flow with power law decay of autocorrelations.  In fact, in the limit $p\to 1$ it can be  shown that the autocorrelation function decays asymptotically as $C(\tau)\sim \tau^{-(\eta-1)}$.

Simulations such as the one presented in the next section show that Eq.~(\ref{eq:anal}) is a good approximation for the order flow decomposition.

\subsection{Alternative causes of persistent order flow between agents}

We want to stress that  a large contribution by $C_{other}$ is a necessary but not sufficient condition for herding.  Herding necessarily involves multiple agents, but the fact that a mechanism involves multiple agents does not mean that it is necessarily due to herding.  For example \cite{Lyons97} has documented the existence of a ``hot potato" dynamic in which dealers trade order flow imbalances among one another. \cite{Yamamoto11} also showed order switching mechanisms could produce correlated action among agents that is similar to herding due to shared information. Such mechanisms could cause a non-zero contribution to $C_{other}$, and might be hard to distinguish from herding.  As we will show in the empirical section, for the LSE data $C_{other}$ is small in comparison to $C_{same}$, so the possibility of alternative mechanisms is a moot point -- the necessary condition for herding is not satisfied.  However the reader should bear in mind that when we refer to ``herding" we are allowing for the possibility of alternative mechanisms that make $C_{other} > 0$.

\section{Models of brokerage\label{brokerageSection}}

As described in the introduction, most of the member firms in the LSE are at least partially acting as brokers, lumping together orders from many investors.  Furthermore, investors may be using multiple brokers, or varying their choice of broker.  Given this, how can we ever hope to make inferences about investors based on data that only identifies exchange members?

In this section we address the distortion caused by observing brokerage codes rather than observing investors directly.  We introduce the concept of a brokerage map, construct two simple examples, and study how the corresponding order flow decomposition is altered by brokerage.
We also study an example in which the social network and brokerage relationship are correlated.

Mirroring the same notation that we used for investors, assume there are $M'$ brokers labeled by an index $i$ and $P^{'i}$ is the trading frequency of broker $i$.  Under the assumption that all the orders submitted by single investors are passed through by brokers unaltered on the same time step, the {\it brokerage map} $B(t):{\mathbb R}^M \to {\mathbb R}^{M'}$ is an $M' \times M$ binary matrix with entries $B_{ij}(t) \in \{0,1\}$, with $\sum_i B_{ij}(t) = 1$.  This means that at each time only one element of each column is different from zero. This element  specifies that agent $j$ gives her order to broker $i$ at time $t$.   Letting $\epsilon^{j}$ be the order flow for single investors and $\epsilon'^{i}$ for brokerages, 
\begin{equation}
\label{brokerageMapDefinition}
\epsilon'^{i}_t = \sum_j B_{ij}(t) \epsilon^j_t.
\end{equation}
The brokerage map $B(t)$ in general depends on time and can be deterministic or stochastic\footnote{The results on the relation between order flow correlation of agents and of brokers could equivalently be obtained by defining the matrix $B$ by assuming that its elements are real numbers, summing to one over columns, and representing the probability that an agent $i$ gives the order to broker $j$.}.  

In the language of signal processing, in the typical case $M' < M$, brokerage reduces the observability of order flow. In a manner similar to Eq.~(\ref{autocdecomp}), the autocorrelation of brokerage order flow can be written in the form
\begin{equation}
C'(\tau) =C'_{same}(\tau)+C'_{other}(\tau).
\label{brokerageAutocdecomp}
\end{equation}
Because of the assumption that brokers do not delay orders or alter signs, $C(\tau) = C'(\tau)$, i.e. the total autocorrelation is unchanged.  In contrast the decomposition $(C_{same}, C_{other})$ suffers a {\it distortion}, i.e. it will be modified to a new decomposition
\begin{equation}
(C_{same}, C_{other}) \stackrel{B}{\rightarrow} (C'_{same}, C'_{other}),
\end{equation}
which depends on the brokerage map $B(t)$ as well as the single investor order flow $\epsilon_i (t)$.  
The goal of this section is to compute the distortion for several simple examples that can then be used as reference cases for interpreting empirically observed brokerage order flow.

We now present some results that will allow us to compute the distortion.  By making use of Eqs.~(\ref{tildep}), (\ref{autocoreDefinition}) and (\ref{brokerageMapDefinition}), we show in the appendix that under the simplifying assumption that $\tilde{P}_{ij} = 0$, the autocorrelation for brokerage order flow can be written
\begin{equation}
C'(\tau)=\sum_{i,j,k,l} B_{ik} (\tau) B_{jl} (\tau) P^{kl} (\tau) C^{kl} (\tau)
\end{equation}
which as before can be split into two components:
\begin{eqnarray}
\label{brokerageComponents}
&&C'_{same}(\tau) = \sum_{i,k,l} B_{ik} B_{il} P^{kl}(\tau)C^{kl}(\tau),\\
\nonumber
&&C'_{other}(\tau) = \sum_{i\ne j, k, l} B_{ik} B_{jl} P^{kl}(\tau)C^{kl}(\tau).
\nonumber
\end{eqnarray}
Let $P'^i$ be the brokerage trading frequency, with $\sum_{i=1}^{M} P'^i = 1$.  From Eq.~(\ref{brokerageMapDefinition}) the relation to the trading frequencies $P^j$ of individual investors is 
\begin{equation}
\label{brokerageTradingFreq}
P'^i = \sum_j B_{ij} P^j.
\end{equation}

We now use these relations to investigate the distortion due to brokerage using several idealized models.  We investigate three different possibilities, ranging from the best case to the worst case.

\subsection{Fixed random brokerage\label{FRB}}

We first begin with the best case.  Under fixed random brokerage (FRB) each investor randomly chooses a single broker and thereafter always executes through that broker only.   This implies that each column of $B_{ij}$ has a single element with value one and all other entries zero, and that $B$ is independent of time.  Under fixed random brokerage the order flow decomposition of the three single investor models presented in the previous section is distorted as follows:
\begin{itemize}
\item
{\it Order splitting.}  The decomposition is unchanged, i.e. $C'_{same} (\tau) = C(\tau)$ and $C_{other} = 0$ and there is no distortion.  This is because for order splitting the off-diagonal elements $P^{i \ne j}(\tau) = C^{i \ne j}(\tau) = 0$, and for  fixed random brokerage the fact that only one entry of any column of $B$ is nonzero implies that $B_{ik} B_{jk} = 0$ for $i \ne  j$.  Thus in Eq.~(\ref{brokerageComponents}) $C'_{other}(\tau) = 0$ and therefore $C_{same}(\tau) = C(\tau)$.  (This is just a consequence of the obvious fact that all the order flow of a single investor splitting orders stays inside that investor's brokerage).
\item
{\it Public information herding model}.  The decomposition is given by Eq.~(\ref{eq:anal}), except that $M$ is replaced by $M'$ and $\mathcal{H} = \mbox{Var}[P]$ is replaced by $\mathcal{H}' = \mbox{Var}[P']$, i.e.
\begin{eqnarray}
\label{eq:anal2}
\nonumber
\frac{C'_{same}(\tau)}{C'(\tau)} & = & \frac{1}{M'}+M' \mathcal{H}',\\
\frac{C'_{other}(\tau)}{C'(\tau)} & = & \frac{M'-1}{M'}-M' \mathcal{H'}.
\end{eqnarray}
This can be seen from the fact that for the public information herding model $P^{ij} = P^iP^j$ and $C'^{ij} = C'(\tau)$.  Using Eq.~(\ref{brokerageComponents}) and Eq.~(\ref{brokerageTradingFreq}), and the definition of $\mathcal{H'}$ (as in Section~\ref{herdingModel1}) gives the result. The size of the distortion thus depends on the reduction in the dimensionality of the space from $M$ to $M'$ as well as the change in the heterogeneity from $\mathcal{H}$ and $\mathcal{H}'$.  The effect of the distortion is to make pure herding at the level of investors acquire a stronger $C_{same}$ component, making it look more like splitting\footnote{
While it might seem surprising that this no longer depends on $M$ or $\mathcal{H}$, one should bear in mind that under fixed random brokerage we are assuming the correlation between order flow at the level of investors is independent of their broker.  Furthermore, $M$ and $\mathcal{H}$ have to be consistent with the set $\{P^{'i}\}$.  Since behavior below the level of the brokerage is not observable, the trading of any investor within a given brokerage has the same effect, and the results are as if there were $M'$ rather than $M$ investors with variance $\mbox{Var[}P'\mbox{]}$}.
\item
{\it Private information herding model.}  We cannot prove the decomposition in this case.  However, as explained below, numerical simulations make it clear that the private information model produces results that are virtually identical to the public information model.
\end{itemize}

\begin{figure}[t]
\begin{center}
\includegraphics[width=0.65\textwidth,angle=0]{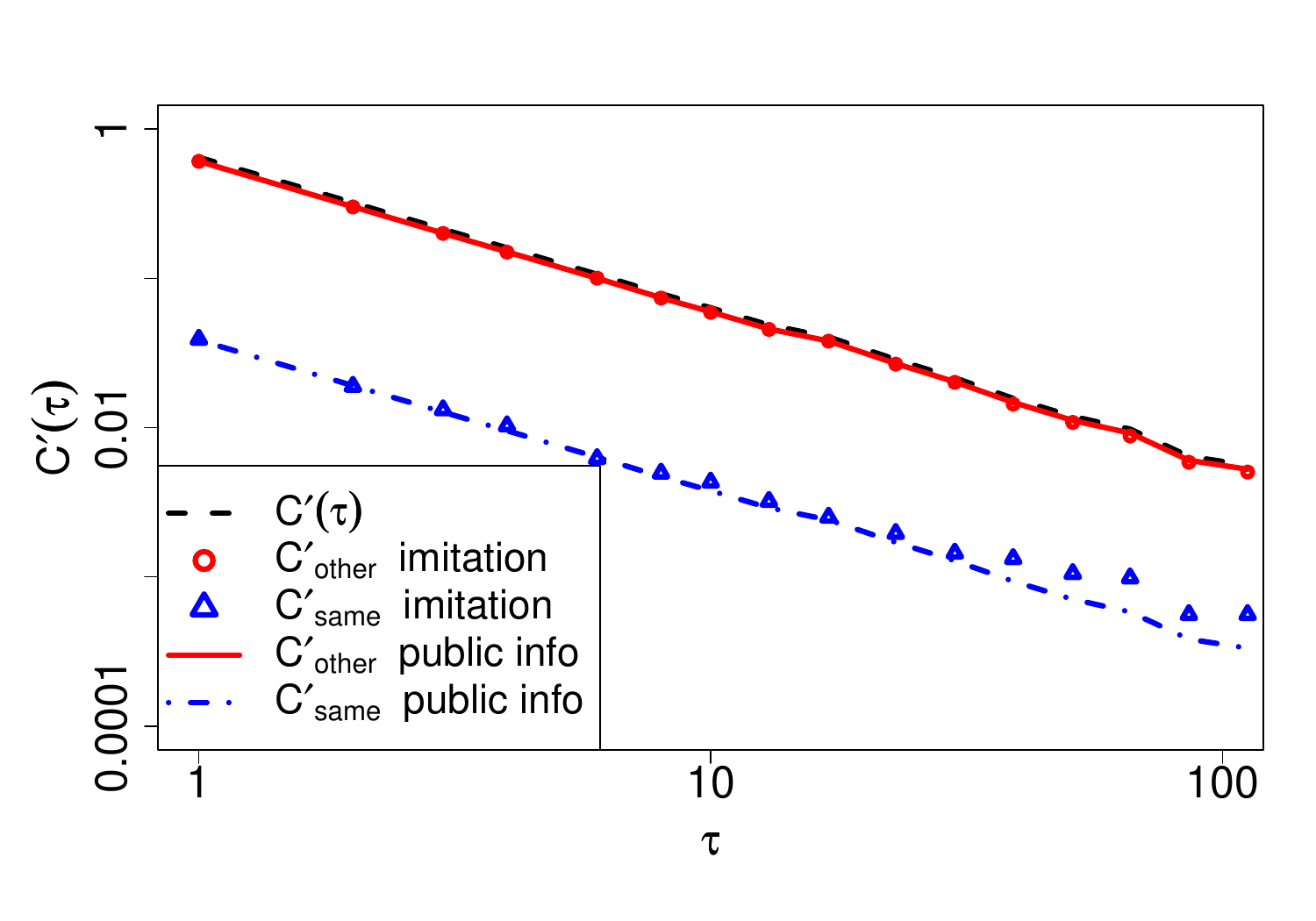}
\caption{\label{fig_network0.1} Decomposition of the autocorrelation of order flow for the two herding models using a fixed random brokerage map.  For the private information (imitation) model, order flow is generated as described in the text, and the splitting (blue triangles) and herding (red circles) contributions  are computed numerically using Eq.~(\ref{components}).  For the public information herding model we assume the same total autocorrelation $C(\tau)$ (black dashed line) as the imitation model, and compute the splitting (blue dots) and herding (red line) contributions using Eq.~(\ref{eq:anal2}).  Broker trading frequencies $P^{'i}$ are chosen to match AZN in the first half of 2009.   Herding strongly dominates splitting and the decomposition for the two models is nearly identical.}
\end{center}
\end{figure}

The private and public information herding models are compared in Figure~\ref{fig_network0.1}, where we calibrate to match AZN during the first half of 2009.  In both cases we construct a fixed random brokerage map  with $M' = 50$ brokers whose trading frequencies $P'^i$ are matched to the 50 largest exchange members for $AZN$ during the first half of 2009.   For the private information model we build an investor social network as described in the previous section with imitation frequency $p=0.9$ and $M = 10,000$ investors, running simulations for $10^6$ time steps.  The decomposition for the public information model is easily computed by calculating $\mathcal{H}'$ from Eq.~(\ref{heterogeneityDefinition}) with $M' = 50$ and plugging into Eq.~(\ref{eq:anal2}).  The resulting decomposition for the two herding models is nearly identical.  In both cases the resulting decomposition is completely dominated by the herding component.  While there is a non-zero splitting component, it is more than an order of magnitude smaller.

Thus for order splitting the fixed random brokerage map introduces no distortion, whereas for herding models the distortion is dependent on the number of brokers and the heterogeneity of their trading frequencies.  The distortion induced for the London Stock Exchange is reasonably small.  Stating this in more quantitative terms, from Eq.~(\ref{eq:anal}) we know that at the level of single investors, providing the order flow is spread across a large number of investors, we expect $S(\tau) = C_{same}(\tau)/C(\tau) \approx 0$.  If we instead observe the same order flow through brokerages, under the fixed random brokerage model, for the LSE we have $M' \approx 80$ and  $\mathcal{H} \approx 10^{-3}$, which according to Eq.~(\ref{eq:anal2}) gives $S'(\tau) = C'_{same}(\tau)/C(\tau) \approx 0.09$.  The distortion is thus only about $9\%$, i.e. the false splitting term is small.

\subsection{Dynamically random brokerage\label{DRB}}

We now investigate the worst case.  To conceal order flow some investors may vary their choice of brokers.  Dynamically random brokerage captures this strategy in its extreme form:  Under dynamical random brokerage, $B_{ij}(t)$ is dynamically random:  On every trade each investor randomly chooses a (typically new) broker according to the broker trading frequencies $P^{'i}$.  There is no memory of the past, and no allegiance of any investor to any broker.  Thus in this case brokerage level order flow carries no information whatsoever about single investors.

What would we observe in this case?  The decomposition is easy to compute:  Random choice implies that $P^{'ij} = P^{'i}P^{'j}$, and therefore the decomposition is once again given by Eq.~(\ref{eq:anal2}).    The resulting decomposition is thus identical to that of the public information herding model under a fixed random broker map.

We have the interesting result that dynamically random brokerage creates the appearance of herding, regardless of what the investors are actually doing.  In contrast to fixed random brokerage, the distortion for herding is small and the distortion for splitting is large.  To make this more explicit, assume the same values for the London Stock Exchange used in the previous section.  If single investors are herding, dynamically random brokerage creates a false non-zero splitting ratio $S'(\tau) \approx 9\%$, the same as that for fixed random brokerage.  But for pure order splitting, the splitting ratio $S(\tau)$ is reduced to $9\%$, in contrast to the correct value of $100\%$ for fixed random brokerage.  This asymmetry is key to allowing us to draw firm conclusions based on brokerage data.


\subsection{Joint model where the social network correlates trading and brokerage\label{correlatedBroker}}

The previous models treat trading and brokerage as independent. What if the trading behavior of investors is correlated with their choice of brokers? To study this question we have generalized the private information herding model to allow for the possibility that neighbors in the influence network also tend to use the same broker.

In the generalized private information model the social network and the brokerage assignments are made in tandem.   As we construct the social network, when we attach each new investor to a previous investor, we assign the same broker with probability $\Phi$ or a random broker with probability $1-\Phi$.  As a result there is a correlation between neighborhood relationships in the trading graph and brokerage assignment\footnote{
An inconvenient feature of this model is that $\Phi$  and $P^{'i}$ are not independent, so it is harder to construct a network with given values of $P^{'i}$.  As $\Phi$ varies from zero to one the trading frequencies of the brokers necessarily become more concentrated.}.
This correlation causes a higher level of distortion for the obvious reason that it means that two investors who frequently mimic each other tend to trade through the same broker, making them look like they are a single investor splitting orders.

For this model we can only compute the distortion numerically; the results are presented in the next section when we compare to empirical data.

\section{Empirical results \label{empiricalResults}}

\subsection{A naive view of the exchange member data}

\begin{figure}[t]
\begin{center}
\includegraphics[width=0.65\textwidth,angle=0]{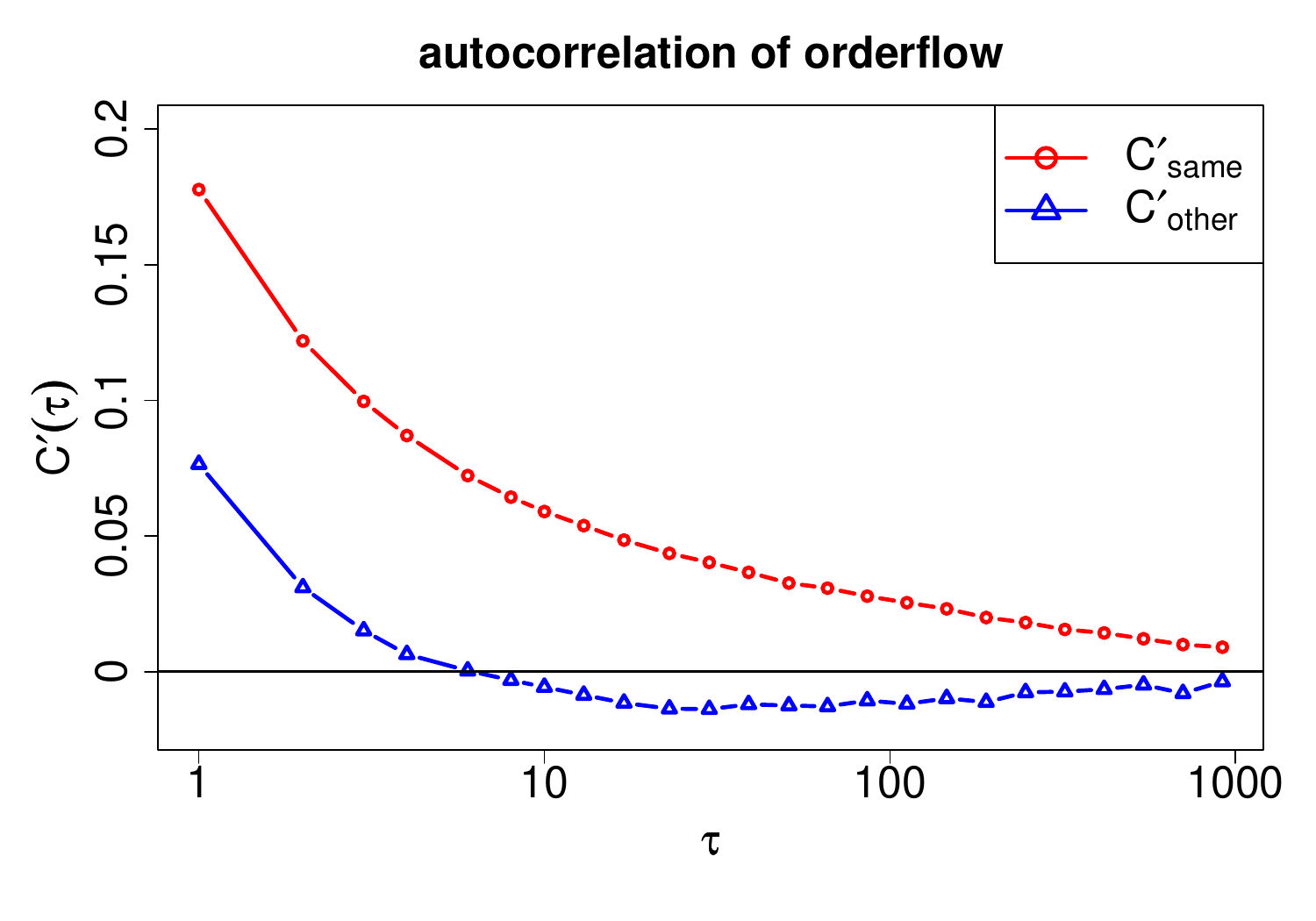}
\end{center}
\caption{\label{splittingDecomp}  A decomposition of the autocorrelation of the order flow into $C'(\tau) = C'_{same}(\tau) + C'_{other}(\tau)$ according to Eq.~(\ref{autocdecomp}) based on exchange membership identifiers.   The horizontal axis is plotted on logarithmic scale and the vertical axis on linear scale.  $C'_{same}(\tau)$ dominates at all lags. Note the negativity of $C'_{other}(\tau)$ for $\tau >10$.  Here we show AZN in the first half of 2009, but all of the 102 samples look very similar.}
\end{figure}

We are now ready to examine the empirical data.  We begin by showing in Figure \ref{splittingDecomp} the decomposition of the autocorrelation of order flow into its components $C'_{same}(\tau)$ and $C'_{other}(\tau)$ according to Eq. \ref{autocdecomp}.  This is based on exchange membership identifiers alone -- in most cases investors are trading through exchange members as brokers, and all we can observe is their behavior aggregated together with others trading through the same broker.

We see that $C'_{same}(\tau)$ is always positive and is substantially larger than $C'_{other}(\tau)$ at all lags.  This is particularly true for larger lags -- by $\tau = 10$, $C'_{other}(\tau)$ is near zero.  

\begin{figure}[ptb]
\begin{center}
\includegraphics[width=0.49\textwidth,angle=0]{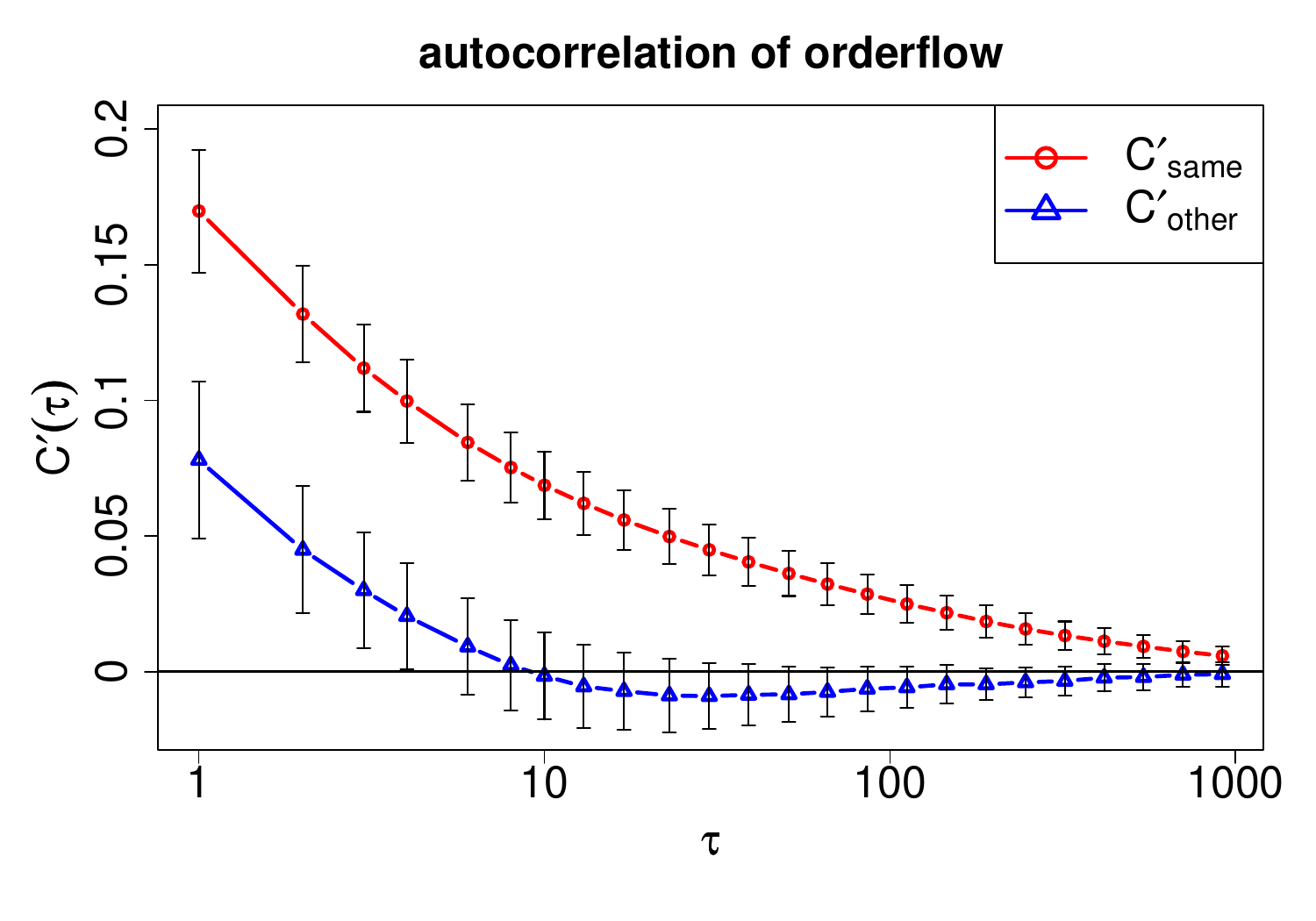}
\includegraphics[width=0.49\textwidth,angle=0]{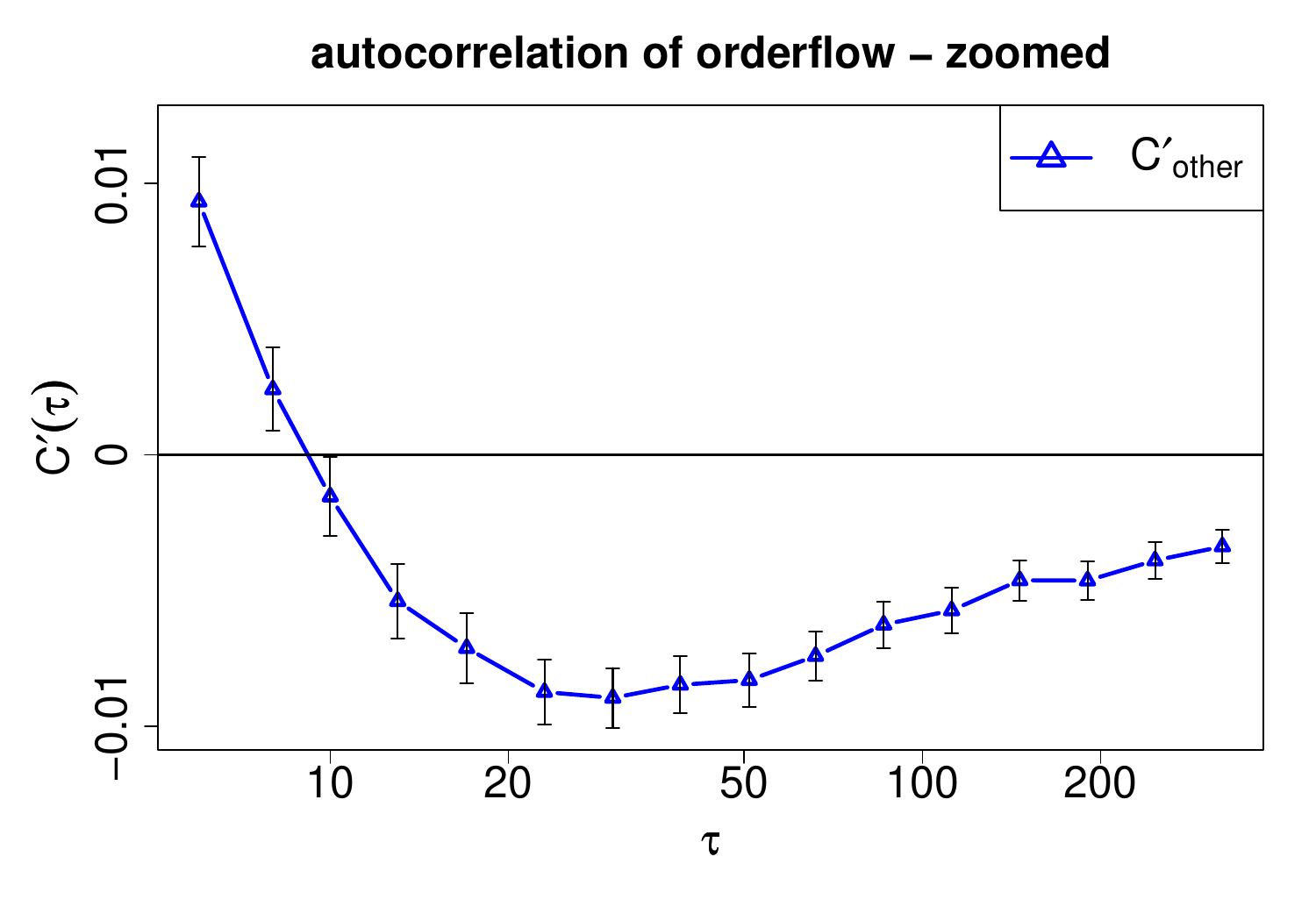}
\end{center}
\caption{\label{splittingherding_ensemble} The decomposition of the autocorrelation for the ensemble of stocks illustrates the consistency with which $C'_{same}(\tau)$ dominates $C'_{other}(\tau)$.  {\it Left panel:} Autocorrelations of market order signs averaged across all 102 samples (spanning nine years and six stocks).  The bars are standard deviations.  {\it Right panel:} $C'_{other}(\tau)$ in the left panel is magnified to better observe its negative contribution. The bars in this panel are standard errors.  For both plots we use logarithmic scale on the horizontal axis and linear scale on the vertical axis.}
\end{figure}

This behavior is remarkably consistent across all 102 samples.   To illustrate this, in the left panel of Figure \ref{splittingherding_ensemble} we plot the decomposition $C'(\tau) = C'_{same}(\tau) + C'_{other}(\tau)$ averaged across all of the 102 samples, and also plot the standard deviation across the samples for each time lag.  There is remarkably little variation across the samples. The standard deviations are small compared to the difference between $C'_{same}$ and $C'_{other}$.  For $\tau\leq 100$, where we have the best statistical reliability, there is not a single case in which $C'_{other}(\tau) > C'_{same}(\tau)$.

The negative value of $C'_{other}$ observed for AZN is not special to this stock or this time period:  Almost all stocks show similar behavior.  To examine this in more detail, in the right panel of Figure~\ref{splittingherding_ensemble} we enlarge the scale and plot only $C'_{other}$.  Instead of showing the standard deviation across the samples, we show the standard error.  The fact that the average value of $C'_{other}$ is consistently negative for $10 \le \tau \le 250$, at many lags by more than three times the  standard error, suggests that this effect is real.   We will return in Section \ref{sec:anti_herd} to perform better statistical tests and shed some light on the cause of this phenomenon.

\subsection{Comparison to null hypotheses:  Can the data be consistent with herding?}

The dominance of $C'_{same}$ over $C'_{other}$ naively suggests that splitting is the dominant cause of the autocorrelation of order flow, and that herding can at best play a minor role.  But is this true?  Or are there circumstances where this behavior could be consistent with herding?  We now make use of the models for investors and for brokerage that we developed in the previous section to try to shed some light on this question by comparing to specific models of investors and brokerage.  By calibrating to the data we can then ask whether the observed behavior can be explained by a given scenario.

\begin{figure}[tb]
\begin{center}
\includegraphics[width=0.65\textwidth,angle=0]{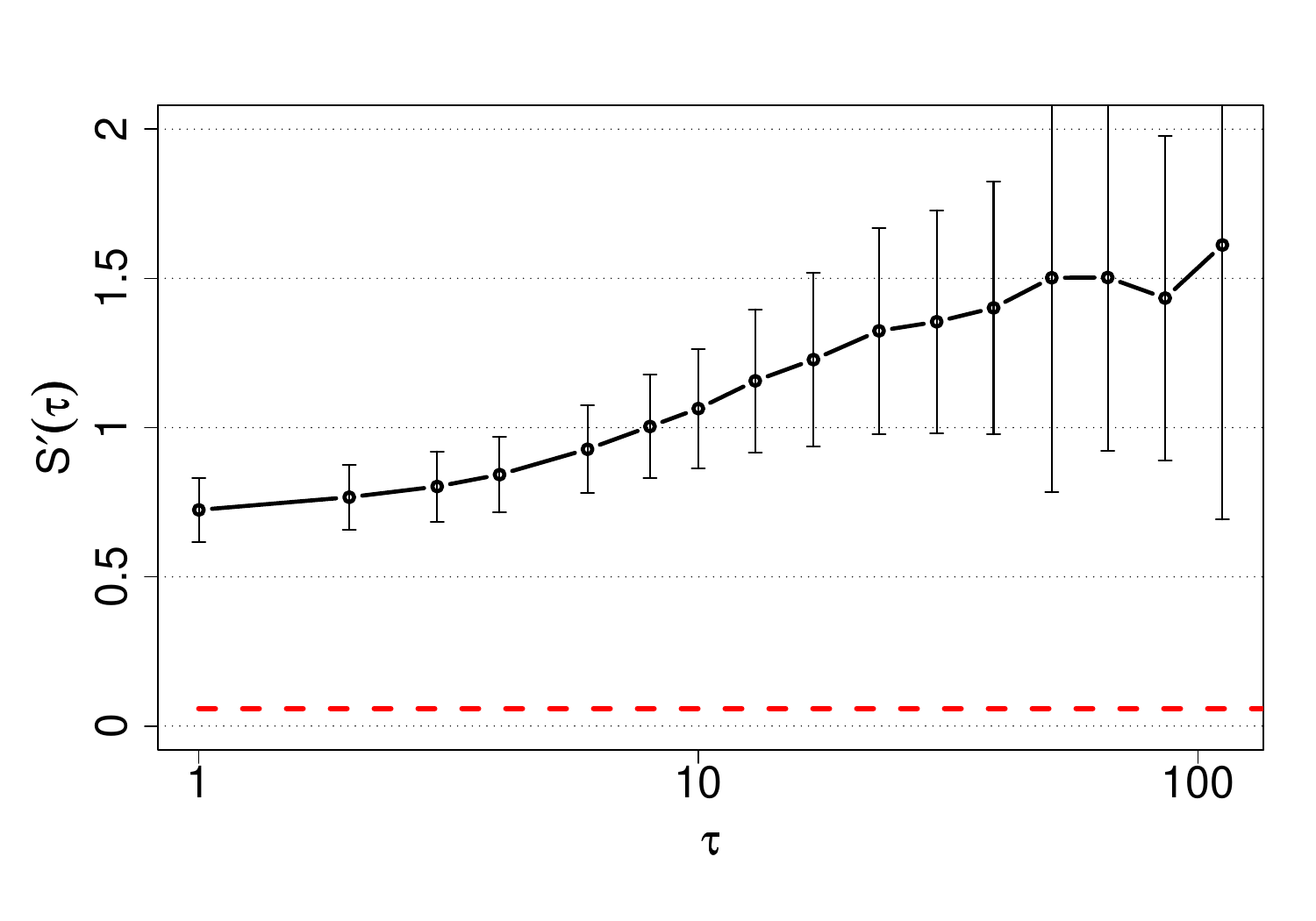}
\end{center}
\caption{\label{sfig} The ratio $S'(\tau) = C'_{same}(\tau)/C'(\tau)$ is compared to a herding null hypothesis.  Results averaged over all 102 samples, corresponding to six different stocks.  The bars represent standard deviations across the 102 samples.  The red dashed line corresponds to the behavior of the herding null hypothesis given in Eq.~(\ref{eq:anal2}) with parameters matched to the data.  The fact that the data are so far from the null hypothesis suggests that herding cannot explain the data.}
\end{figure}

The clarity of the answer is aided by the ubiquity of Eq.\ (\ref{eq:anal2}).  As we saw in the previous section, this corresponds to the following scenarios:
\begin{itemize}
\item
{\it Either herding model/fixed random brokerage}.
\item
{\it Any investor model/dynamic random brokerage}.
\end{itemize}
If we calibrate Eq.\ (\ref{eq:anal2}) to the data using the measured values of $M'$ and $\mathcal{H}' = \mbox{Var} [P']$, then it predicts the decomposition we should observe under any of these scenarios. 

\subsubsection{Dependence on lags}

In  Figure~\ref{sfig} we plot $S'(\tau) = C'_{same}(\tau)/C'(\tau)$ as a function of $\tau$.  Note that $S'(\tau)$ is larger than one if $C'_{other}$ is negative.  We show the average result for all 102 samples.
For $\tau = 1$, $C'_{same}(\tau)$ accounts for about $75\%$ of the autocorrelation and $C'_{other}(\tau)$ explains about $25\%$. For larger lags $C'_{same}$ becomes relatively even more important. For lags larger than roughly 10, $S'(\tau)$ becomes larger than one because $C'_{other}(\tau)$ becomes negative, and $S'(\tau)$ rises to about $1.5$.  This means that $C'_{same}(\tau)$ is about 2.5 times the absolute value of the $C'_{other}(\tau)$.  For large $\tau$ the standard deviation across the samples becomes large because $C'_{same}$ and particularly $C'_{other}$ are small, and we are taking ratios of small numbers.

This is compared to the null hypothesis of Eq.~(\ref{eq:anal2}), which is shown as a dashed red line.  For the data $S'(\tau) = C'_{same}(\tau)/C'(\tau)$ increases from about $0.7$ to $2$ with increasing $\tau$, in contrast to the predicted value, which is roughly $0.06$ independent of $\tau$.  Thus $C'_{same}$ is a factor of $10-30$ higher than predicted under the null hypothesis.

\subsubsection{Dependence on heterogeneity}

\begin{figure}[t]
\begin{center}
\includegraphics[width=0.65\textwidth,angle=0]{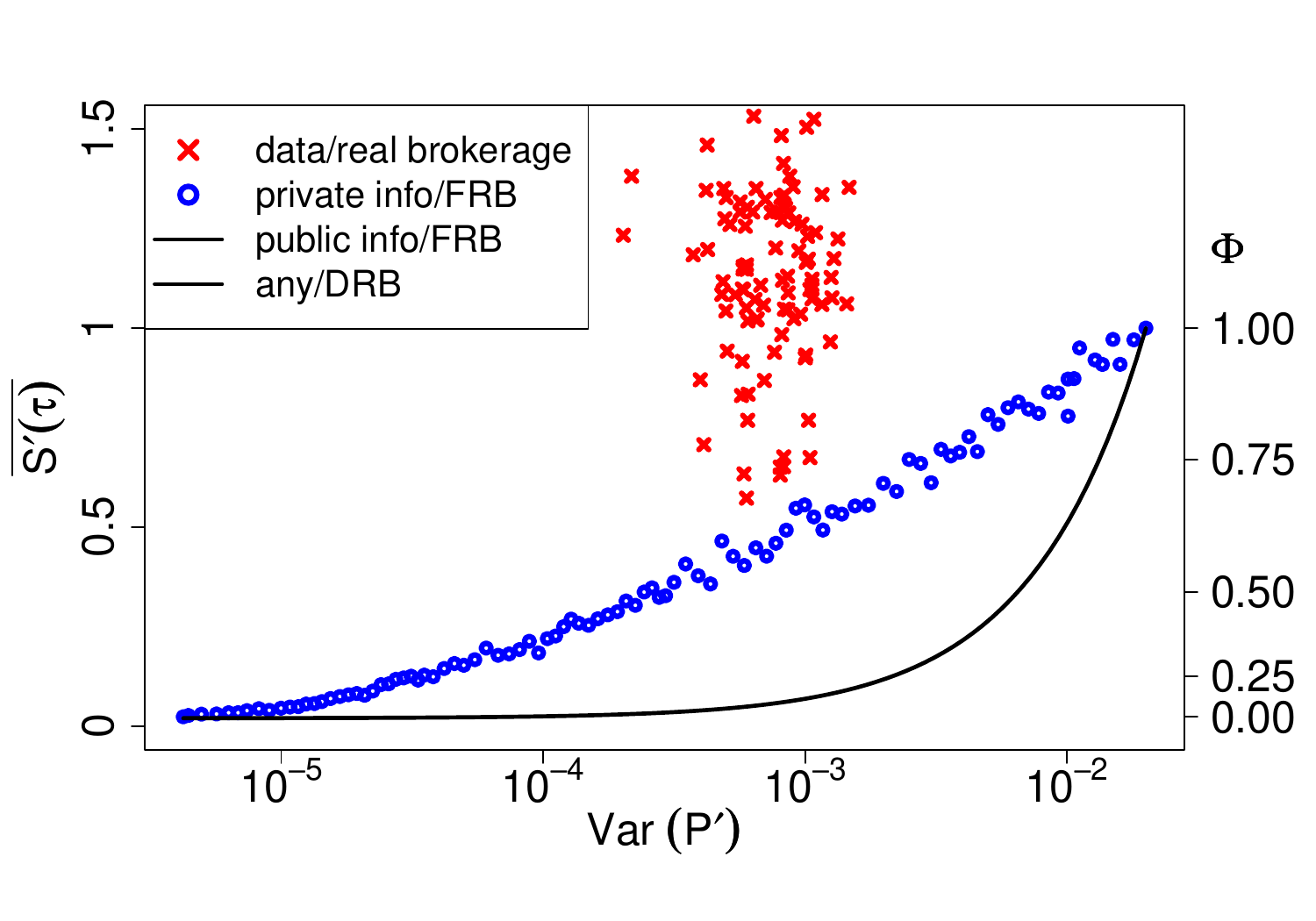}
\caption{\label{networkBrokerage}  Comparison of the null hypotheses for investor behavior and brokerage maps to real data.  The left vertical axis indicates the ratio $S'(\tau) = C'_{same}(\tau)/C'(\tau)$ averaged over $1\le\tau\le 50$ and the horizontal axis is the heterogeneity parameter $\mathcal{H}' = \mbox{Var}[P']$, corresponding to the variance of the brokerage trading frequencies.  On the left the trading frequencies are equal and thus the heterogeneity $\mathcal{H}'$ is small, and on the right trading is concentrated in a few brokerages and the variance is large.  The $102$ data samples (indicated by red x's) are compared to the predictions of three different null hypotheses.  The null hypotheses are generated by combining a model for investor herding with a model of brokerage.  The black curve is the public information model and Fixed Random Brokerage; alternatively it represents the prediction of the Dynamically Random Brokerage (DRB) model under any investor model.  The blue circles represent simulations of the private information herding model with Fixed Random Brokerage (FRB), with the correlation $\Phi$ between the brokerage choice and trading signal shown on the right vertical axis. Under any brokerage model all $102$ samples are outside the range predicted by any of the null hypotheses that involve herding.}
\end{center}
\end{figure}

As we have seen, the heterogeneity parameter $\mathcal{H}'$ plays a key role in determining the degree of distortion due to brokerage.  In Figure \ref{networkBrokerage} we compare the data to the null hypotheses by plotting an average of $S'(\tau) = C'_{same}(\tau)/C'(\tau)$ for lags $1\le\tau\le 50$ and plotting this against the heterogeneity parameter $\mathcal{H}'$.  This figure contains several key results and deserves some detailed discussion.

The black curve indicates Eq.~(\ref{eq:anal2}), which as stressed above represents several different scenarios that can be used as null hypotheses.  $S'(\tau)$ remains close to zero for small $\mathcal{H}'$, but rises steeply from zero to one when $\mathcal{H}' > 10^{-3}$.   This is because as $\mathcal{H}'$ the distortion increases and gets large $C'_{same}$ becomes large.  Note that the data have $\mathcal{H}'  \simeq 10^{-3}$, indicating that we are in a region where the distortion is still small.

The blue circles indicate the generalized private information herding model with fixed random brokerage, in which the choice of brokerage is correlated to that of neighbors in the social network.  In other words, this represents a worst case scenario in which investors that tend to imitate each other also tend to choose the same broker.    In this case we also plot the correlation parameter $\Phi$ on the vertical axis on the right.  (Since the relation between $\Phi$ and $S$ is nonlinear, this has a nonlinear scale).   The value of $S$ under this null hypothesis increases roughly linearly as a function of the heterogeneity parameter $\mathcal{H}'$, and due to the correlation, this is a stricter null hypothesis.

The value for each data sample is represented by a red ``x" in Figure \ref{networkBrokerage}.  The data samples are well-separated from the null hypotheses in all $102$ cases, indicating that none of them can explain the data.  The only null hypothesis that comes close is the one that correlates order placement under private information with broker choice.  A few of the samples are close to this null hypothesis, but most of them have values of $S$ that are a factor of two or three greater.  Note also that in order to even come close, for these samples it is necessary to assume that the correlation between trading and broker choice is the order of $60\%$, which seems high.


To summarize, the results in this section make it clear that we can strongly reject both (1) herding with fixed random brokerage (even when imitation and brokerage choice are correlated) and (2) dynamically random brokerage.  In a certain sense these bracket the way in which investors can choose brokers.  In case (1) the investors are highly faithful to their brokers, in case (2) they are not faithful at all.  We are able to reject (2) because dynamically random brokerage looks indistinguishable from herding.  Thus if we had observed a large correlation $C'_{other}$, we would not have been able to reject this possibility.  The inference problem is highly asymmetric:  Splitting leaves a clear signal that is not distorted by fixed random brokerage, and at the same time is easy to distinguish from dynamical random brokerage.

Our analysis allows us to make inferences about the nature of brokerage choice, which is interesting in and of itself.  The fact that the dynamically random brokerage model is so strongly rejected makes it clear that investors do not randomize their brokerage choice.  In order to observe the large values of $S'(\tau)$ seen here it must be the case that they must have a substantial degree of consistency in choosing brokers, and that the typical investor only uses a small number of brokers.



\subsection{Can the splitting scenario explain the data?}

We have seen that the herding null hypotheses are rejected by the data, but what about the splitting null hypothesis?  As already discussed in the introduction, the order splitting model of \cite{Lillo05b} is capable of matching the empirical autocorrelation $C'(\tau)$ quite well.  The splitting ratio of this model is $S'(\tau) = 1$, independent of $\tau$ and independent of the brokerage map.  This ratio is somewhat too high for $\tau \le 5$, where $S \approx 0.7$.  This could be due to one of two reasons, that we cannot distinguish:  (1) Some short term herding behavior or (2) partially dynamic brokerage choice\footnote{
By partially dynamic brokerage choice we mean that some investors (but not others) might randomly vary their brokers, or investors might have partial allegiance to a small set of brokers, randomly choosing between them but avoiding the others.  As the magnitude of the random choice increases so does the herding component, and as we know from the previous scenario, unless a single brokerage dominates the trading (which is not the case for the real data), this strongly distorts the decomposition toward $C_{other}$.}.
For $\tau \ge 10$ the splitting ratio $S'(\tau) = 1$ predicted under the order splitting scenario is actually too low, due to the negative contribution of the herding component.   Thus, the order splitting model explains the basic fact of the dominance of order splitting, but to get a better match to the data one would need to include other effects, such as a small component of herding, allowing the possibility of negative contributions to herding, and partially dynamic brokerage (i.e. allowing investors to split their trading across a few brokers, which gives the appearance of herding).

\subsection{The evolution of splitting versus herding through the decade}

During the last decade there has been an enormous rise in the use of algorithmic brokerages, which take large institutional orders, split them into small pieces and execute them incrementally.  One might expect this activity to be reflected in an increase in order splitting over the course of our ten year data set, which covers 2000 - 2009.

\begin{figure}[tb]
\begin{center}
\includegraphics[width=0.65\textwidth,angle=0]{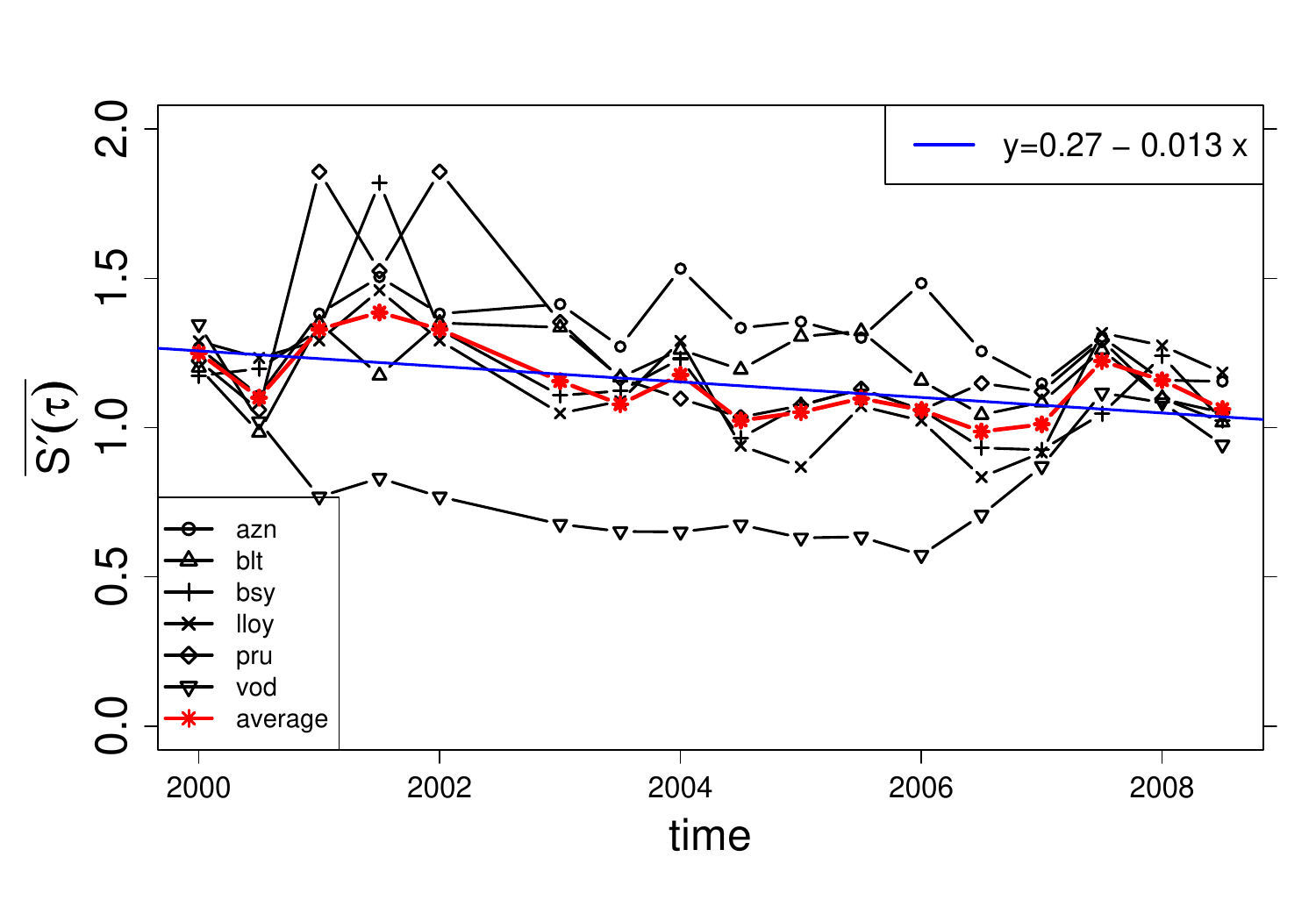}
\end{center}
\caption{\label{splittingEvolution} The evolution of order splitting versus herding throughout the decade.  For each six month period we compute the splitting ratio $S'(\tau) = C'_{same}(\tau)/C'(\tau)$ averaged over $1\leq\tau\leq 50$ for each of the stocks in our data set.  The red line with stars shows the average of the splitting ratios in each period, and the blue solid line is a linear regression.  The regression is negative, suggesting the dominance of splitting over herding is actually decreasing slightly through time (and in any case is clearly not increasing).}
\end{figure}

To investigate this we compute the splitting ratio $S'(\tau) = C'_{same}(\tau)/C'(\tau)$ for each of the stocks in our data set in each half-year.  The result is shown in Figure~\ref{splittingEvolution}.

The surprising result is that we observe no increase in the average splitting ratio with time -- in fact, a linear regression indicates a slight decrease.  The only possible exception is Vodafone,  which shows an increase from 2006-2009 (but experienced a substantial decrease during the first two years).  These results indicate that, at least in comparison to herding, order splitting was a common activity even before the widespread use of algorithmic trading, and that the rise of algorithmic trading has been a relatively small effect.

\subsection{Heterogeneity of individual members}

In this section we study the behavior of individual exchange members and show that their behavior is remarkably consistent, with a few exceptions.  The decomposition of Eq.~(\ref{ptilde}) makes it clear that persistence in order flow is driven by two factors:  persistence in order sign, measured by $C^{ij}(\tau)$, and persistence in activity, measured by $P^{ij}(\tau)$.  Because they appear multiplicatively, both factors are needed to get a large autocorrelation.  Insofar as splitting is dominant over herding the diagonal terms $C^{ii}(\tau)$ and $P^{ii}(\tau)$ will dominate the contribution due to the off-diagonal terms.

\begin{figure}[ptb]
\begin{center}
\includegraphics[width=0.49\textwidth,angle=0]{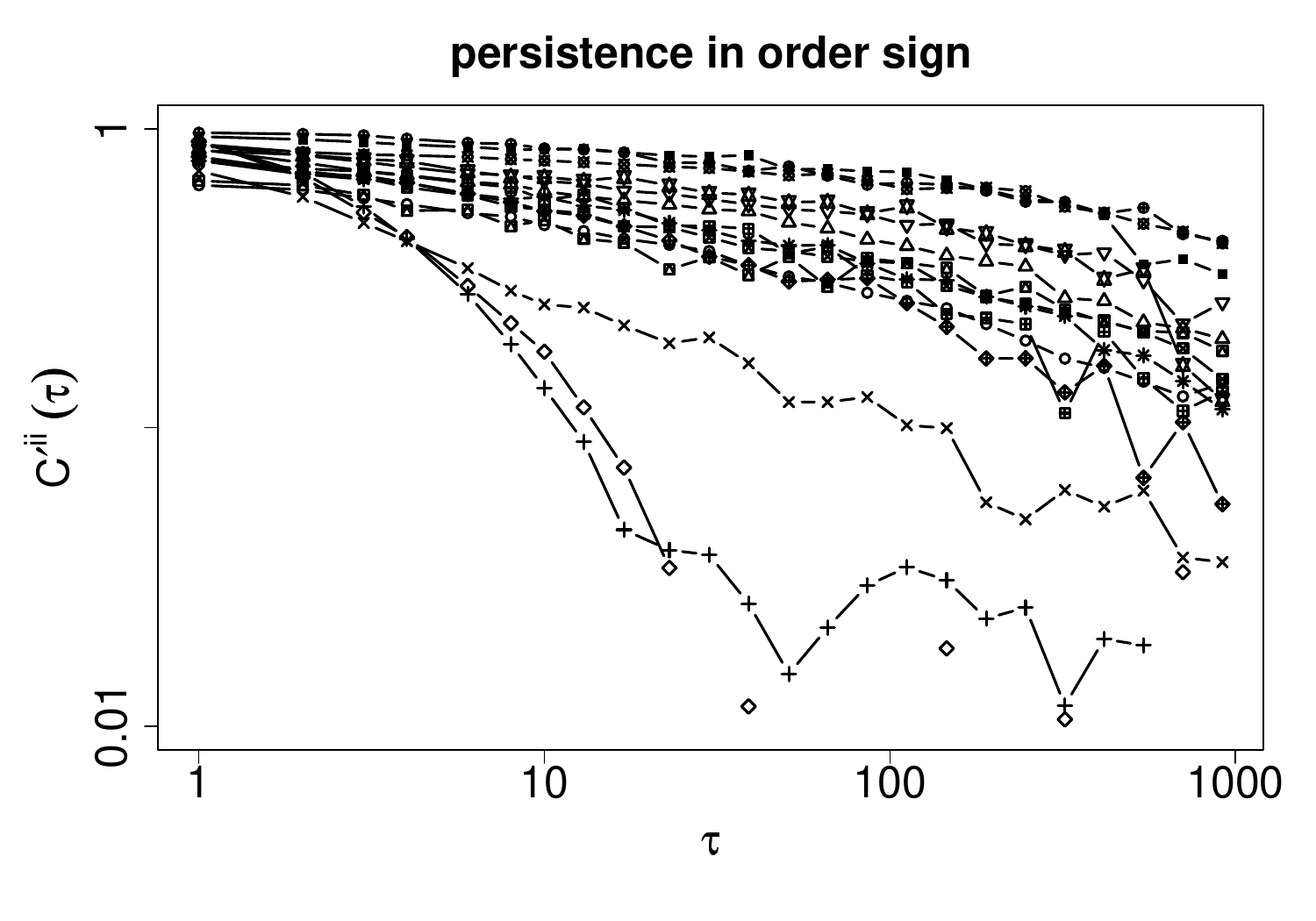}
\includegraphics[width=0.49\textwidth,angle=0]{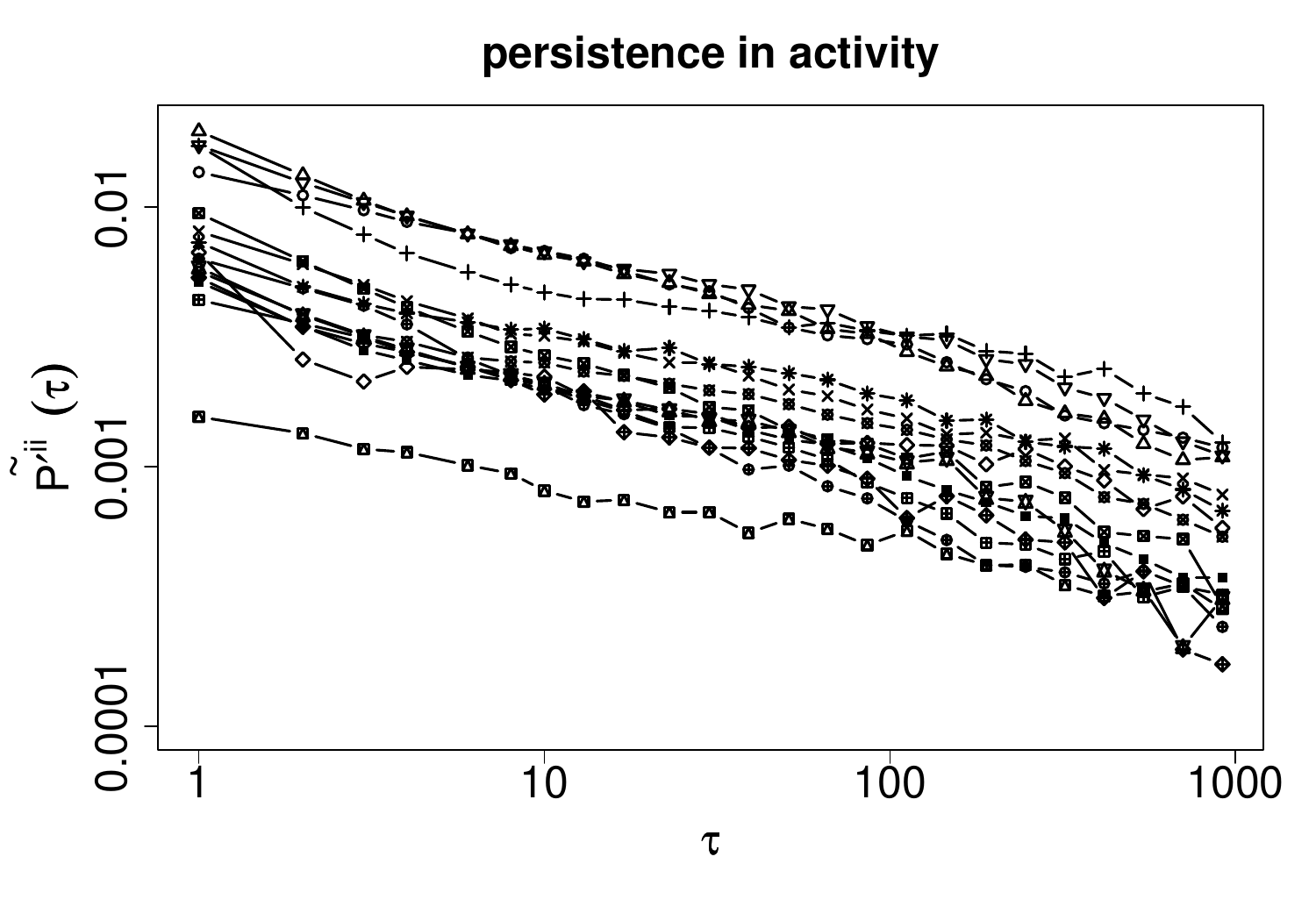}
\end{center}
\caption{\label{splittingind} Heterogeneity of the contribution of individual exchange members (labeled by $i$) in the diagonal component of the autocorrelation of order flow.   Following the decomposition of Eq.~(\ref{ptilde}), on the left we plot $C'^{ii}(\tau)$, which measures persistence in trading direction and on the right we plot $\tilde{P}'^{ii}(\tau)$, which measures persistence in activity.  We show data for the 15 most active members for AZN for the first half of 2009, using distinctive symbols to indicate each member (consistently in both plots). For one of the members (labeled by diamonds) $C'^{ii}(\tau)$ decays more quickly than the others and for some values of $\tau > 10$ is negative, hence the diamonds cannot be plotted on logarithmic scale and cannot be connected.   This member is the fifth most active.
}
\end{figure}

To understand how these vary across the members of the exchange, in Figure~\ref{splittingind} we show $C'^{ii}(\tau)$ and $\tilde{P}'^{ii}(\tau)$ for the 15 most active members.

The patterns for trading sign are consistent in most cases.  $C'^{ii}(\tau)$ decays very slowly, but the curves tend to steepen slightly with increasing $\tau$.  There are three members for which  $C'^{ii}(\tau)$ decays  faster than the others, and two members in particular for which the behavior is dramatically different, suggesting that these members follow a different business model.

\begin{figure}[tb]
\begin{center}
\includegraphics[width=0.65\textwidth,angle=0]{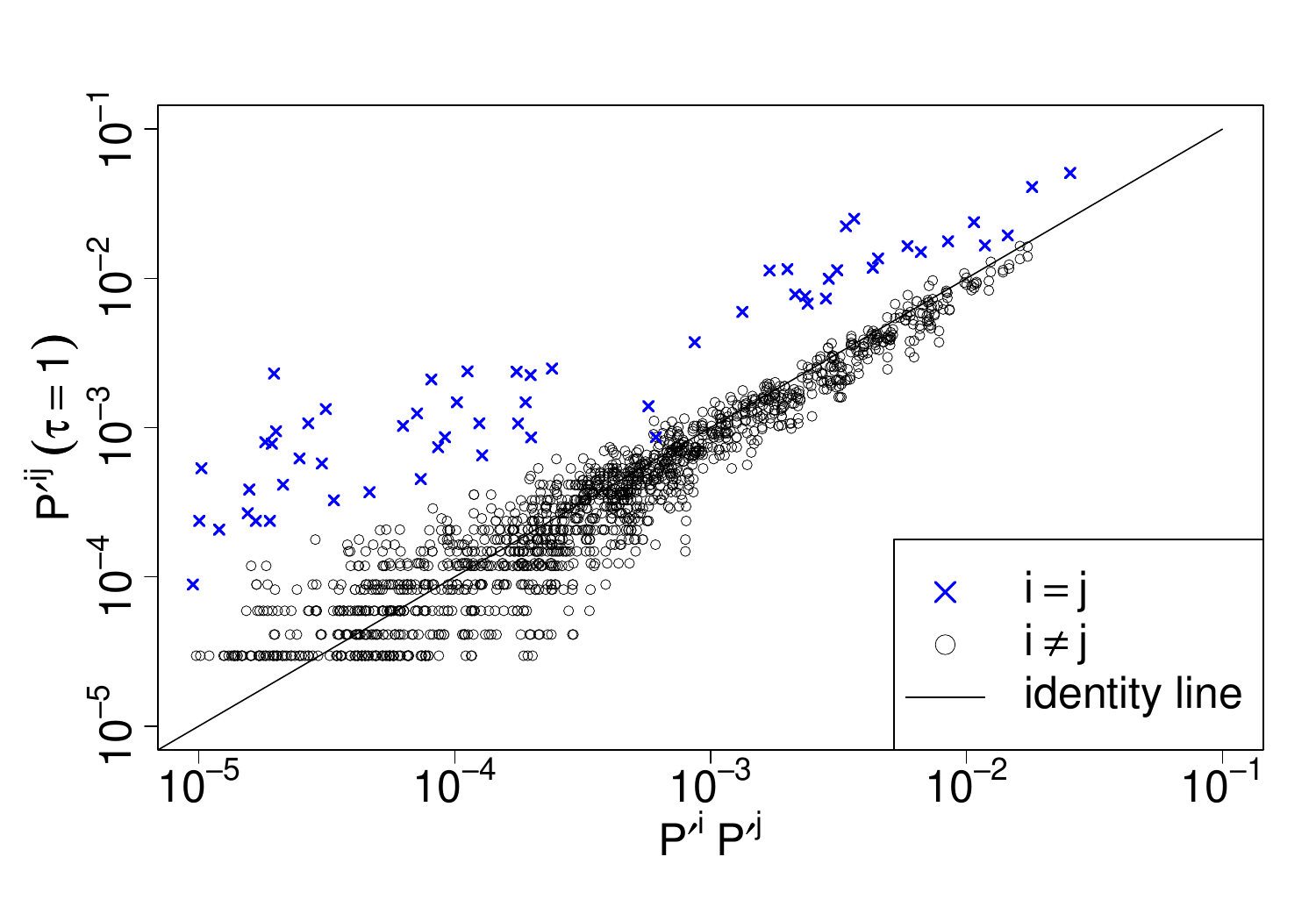}
\end{center}
\caption{\label{Pratio} Persistence in activity.  The joint probability of activity $P'^{ij}(\tau = 1)$ as a function of $P'^{i}P'^{j}$, the hypothetical joint probability under the assumption of independence.   Each symbol indicates a given pair of members $i$ and $j$ for AZN during the first half of 2009.  Diagonal elements $i=j$ are indicated by blue crosses and off-diagonal elements $i\neq j$ by black circles.  The plot is on double logarithmic scale.  For comparison the identity line $P'^{ij}(\tau)=P'^{i}P'^{j}$ is shown as a solid black line.  The fact that the diagonal elements $P'^{ii}(\tau)$ consistently lie well above the identity line is consistent with the large contribution of splitting.  In contrast the off-diagonal elements $P'^{ij}$ (with $i \ne j$) tend to cluster along or even slightly below the identity line.}
\end{figure}

The integral of $\tilde{P}'^{ii}(\tau)$ can be seen as a measure of the activity clustering, thus large values indicate that when the corresponding member becomes active, it will likely persist in this state.
The right panel of Figure~\ref{splittingind} shows that for $\tilde{P}'^{ii}(\tau)$ all of the members show strikingly similar behavior; the individual curves are fairly straight and parallel to each other.  Since we are plotting the data on double logarithmic scale, the straightness of the lines indicates that the
individual exchange members are all reasonably well approximated as a power law, and the fact that the lines are parallel indicates an exponent that is independent of the exchange member.  The offsets indicate differences in scale.  Thus from the point of view of  patterns of trading activity, all members behave in more or less the same manner. This indicates that the activity clustering of all members behaves similarly except for scale.  The scale varies by nearly an order of magnitude, and is strongly correlated with the trading volume of the member.  The Spearman rho of $P'^i$, which measures the trading volume, and $\tilde{P}'^{ii}(1)$, which measures the persistence, is typically very high, for example $0.92$ for AZN in the first half of 2009.  Because $(P'^i)^2$ is subtracted in computing $\tilde{P}'^{ii}$, such a large correlation is not automatic.  This indicates that the most active members also tend to trade in a more clustered manner than the less active firms, i.e. if they become active, they tend to remain active, and vice versa.  In contrast $C^{ii}$ and $P^i$ are essentially uncorrelated, as are $C'^{ii}(1)$ and $\tilde{P}'^{ii}(1)$, indicating that trading direction and trading activity are uncorrelated.

A different method of visualizing the heterogeneity of trading activity is presented in Figure~\ref{Pratio}. We plot $P'^{ij}(\tau = 1)$ against $P'^{i}P'^{j}$ for all pairs of members $i$ and $j$.  If there is no coordination between the activity of different brokers, we expect the off-diagonal elements $i \ne j$ (shown as black circles) to be close to the identity line.  This is indeed precisely what is observed; the off-diagonal elements are somewhat below the identity line, corresponding to the negative contribution of herding.  The diagonal elements $i = j$, in contrast, are well above the identity (shown as blue x's). This shows that there is strong persistence in the activity of brokers, but little coordination between the activities of different brokers.  Larger values of $\tau$ show similar behavior though with somewhat smaller amplitude.

These results indicate that the persistence of order flow stems from the persistence in both trading direction and activity of individual members of the exchange.  They also show that the dominance of splitting over herding is also apparent in the activity level $P'^{ij}$, where we see that the diagonal elements $P'^{ii}$ dominate over the off-diagonal elements $P'^{ij}$ with $i \ne j$.

\section{Clues about the cause of the negative contribution of herding\label{antiHerding}}
\label{sec:anti_herd}

As we have shown in Section~\ref{empiricalResults}, the herding component of the order flow autocorrelation is often negative for $\tau > 10$.  This implies that buying by one investor tends to invoke subsequent selling by other investors.  In this section we first test the statistical significance of this phenomenon and then perform some empirical investigations that give a clue as to its origin.

\subsection{The negative contribution of herding is statistically significant}

\begin{figure}[tb]
\begin{center}
\includegraphics[width=0.65\textwidth,angle=0]{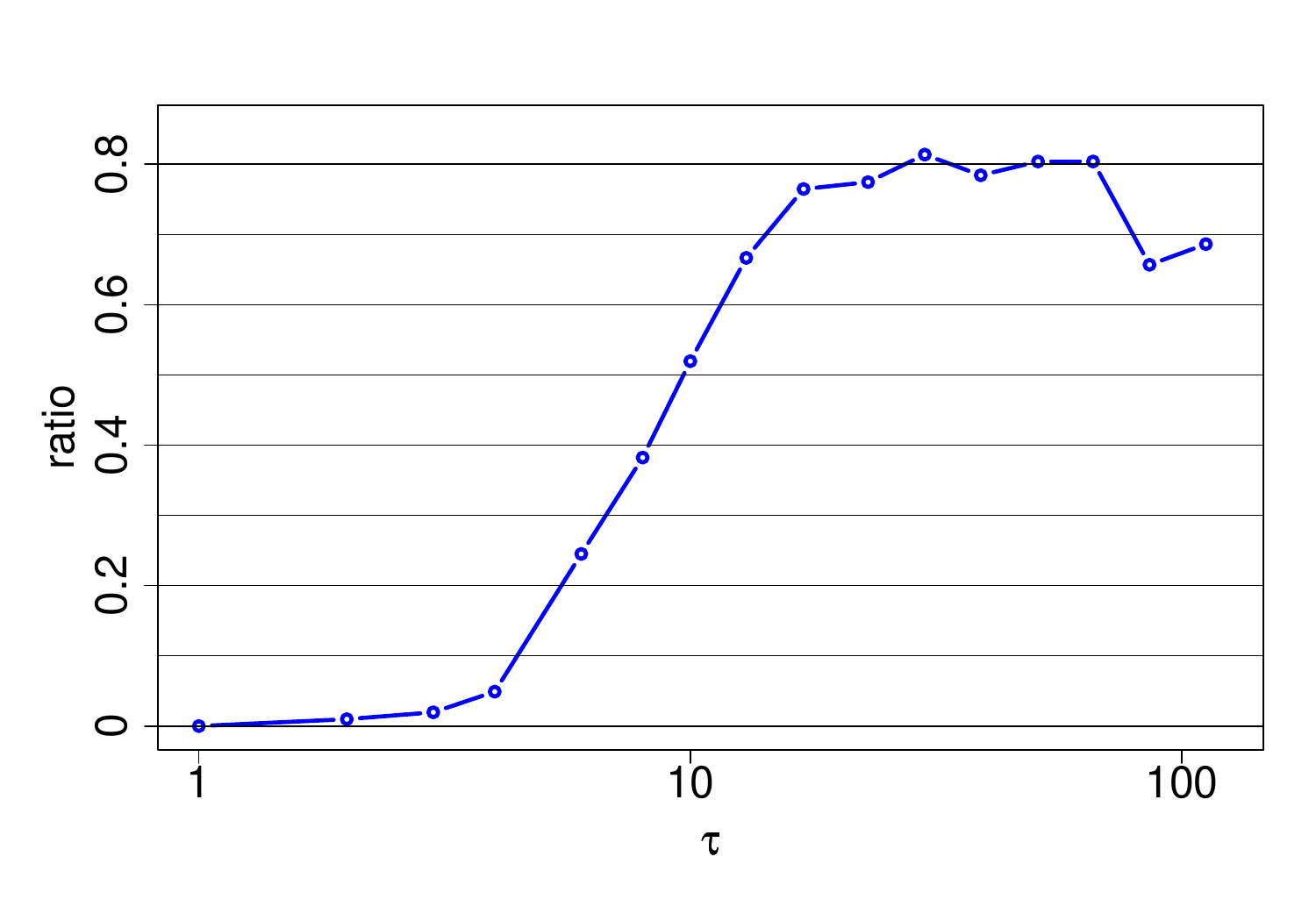}%
\end{center}
\caption{\label{test} Statistical significance of the negative contribution of herding.  The vertical axis plots the fraction of the 102 data sets for which the herding component of the autocorrelation $C'_{other}$ has a $p$ value less than $5\%$ under the IID null hypothesis as a function of the lag $\tau$.}
\end{figure}

To assess whether the negative contribution of herding is statistically significant we perform a one-sided significance test by comparing the real data to the null hypothesis that both the order signs and brokerage codes are assigned randomly.  Realizations of this null hypothesis are obtained by randomly shuffling both the signs and brokerage codes.  For each of the 102 datasets we produced $10^6$  realizations of the null hypothesis, and for each realization we computed the splitting and herding components of the autocorrelation.
Then for each lag $\tau$ we estimated the fraction of random realizations having a herding component smaller than the value observed in the corresponding real sample. If this fraction is smaller than $5\%$, we reject the one-sided null hypothesis. Figure \ref{test} shows the fraction of sets for which we reject the null hypothesis. For small values of $\tau$ we never reject the hypothesis because the real herding component of the autocorrelation function is significantly positive -- this is expected since we do not observe negative contributions in this region. For values of $\tau$ between roughly $15$ and $80$ we reject the null hypothesis in approximately $80\%$ of the sets. This means that for these lags the negativity of the herding component of the autocorrelation is statistically significant. We now explore the possible origin of this phenomenon.

\subsection{What underlies the negative contribution of herding?}

We now show that the negative contribution of herding is associated with a difference in the response of brokers to market orders, depending on whether or not the order changes the price and whether or not it is from the same or a different broker. In particular we find that if a market order placed by broker {\it i} changes the price, broker {\it j} is less likely to place market orders in the same direction, while the behavior of broker $i$ is unchanged\footnote{
The notion that whether or not a market order changes the price is an important factor was inspired in part by a previous study by \cite{Eisler09}; see also T\'oth et al. (\citeyear{Toth11}).}.

\begin{figure}[tb]
\begin{center}
\includegraphics[width=0.49\textwidth,angle=0]{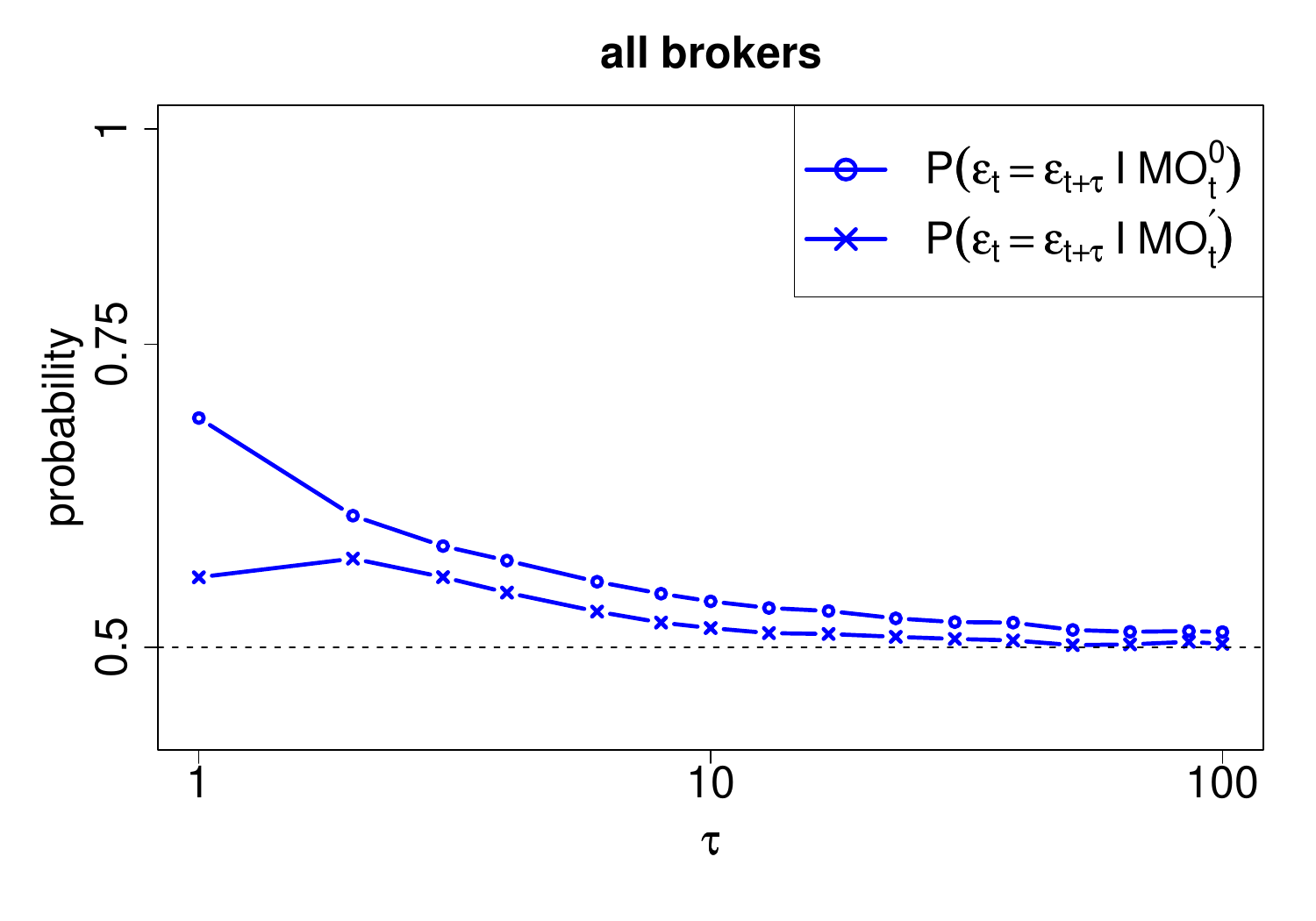}
\includegraphics[width=0.49\textwidth,angle=0]{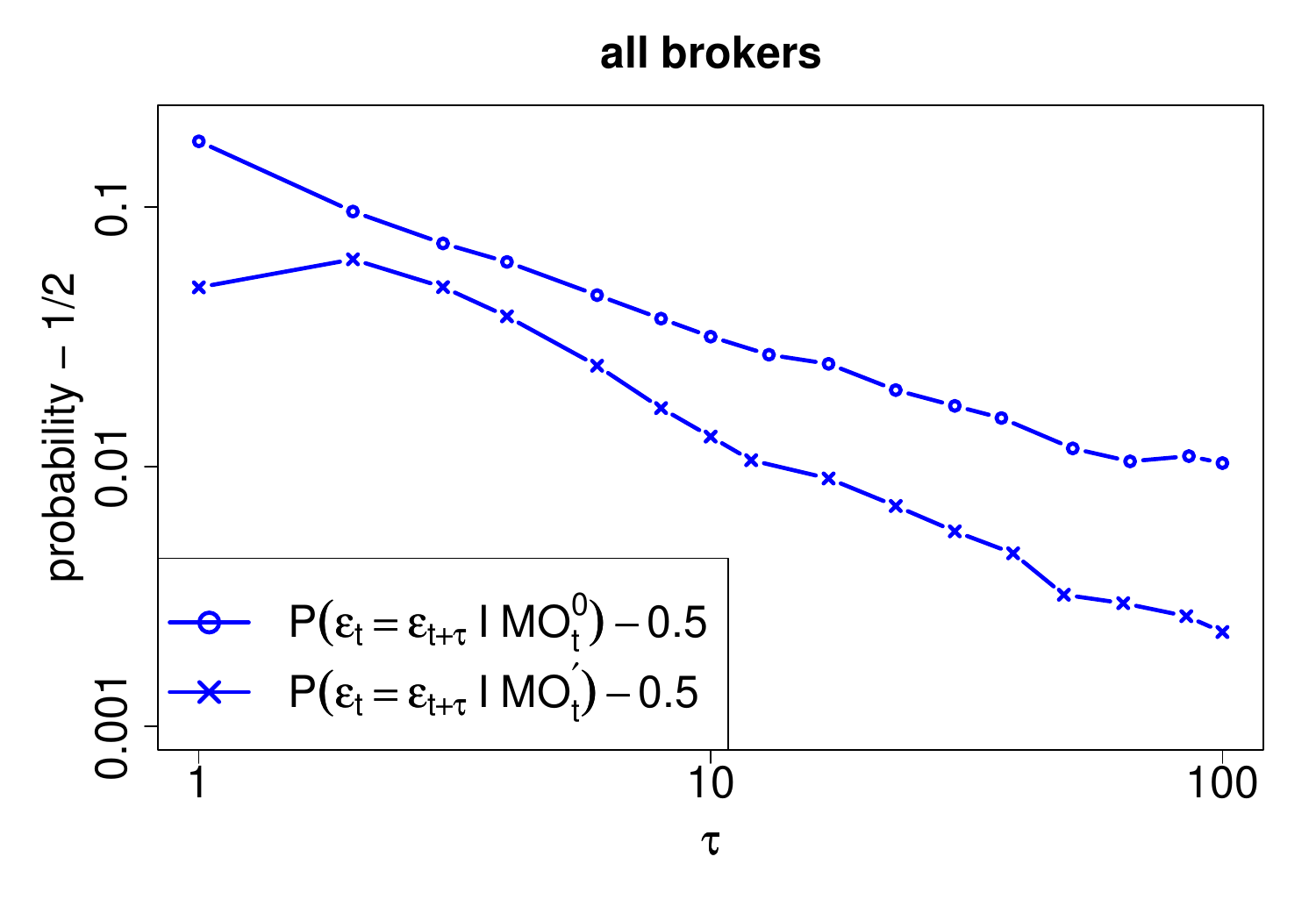}
\end{center}
\caption{\label{same} Probability that market orders at times $t$ and $t+\tau$ have the same sign, conditioned on whether or not the first market order changed the price. The case of no price change is given by $P(\epsilon_t=\epsilon_{t+\tau}|MO_t^0)$ (represented by circles) and the case of a price change by $P(\epsilon_t=\epsilon_{t+\tau}|MO_t')$ (represented by "x").
The plots on the left and right are the same, except that on the right we subtract the unconditional probability $P(\epsilon_t) = 1/2$ to make the functional dependence on time and the difference in the convergence rates clearer.  When the original market order changes the price, the decay is much faster.  Data is for AZN for 2000-2009.}
\end{figure}

We use the notation $MO_t^0$ for a market order at time $t$ that does not change the price and $MO_t'$ for a market order that changes the price.  Conditioned on either of these events, the probabilities for subsequent market orders to have the same sign are
\begin{eqnarray}
P(\epsilon_t=\epsilon_{t+\tau}~|~MO_t^0)\nonumber\\
P(\epsilon_t=\epsilon_{t+\tau}~|~MO_t')
\label{prob1}
\end{eqnarray}
Assuming that on average buy and sell trades are equally likely and that brokers' inventories are bounded, these probabilities for large $\tau$  should converge to the unconditional probability $P(\epsilon_t) = 1/2$. In Figure \ref{same} these are shown for AZN, plotted as a function of $\tau$ on a double logarithmic scale.  To make the $\tau$ dependence clearer, we also plot the excess probabilities
\begin{eqnarray*}
\tilde{P}^0(\tau) \equiv P(\epsilon_t=\epsilon_{t+\tau}~|~MO_t^0) - 1/2\\
\tilde{P}'(\tau) \equiv P(\epsilon_t=\epsilon_{t+\tau}~|~MO_t') - 1/2.
\end{eqnarray*}
The decay of both $\tilde{P}^0$ and $\tilde{P'}$ is approximately a power law.  However, when the original market order changes the price, the decay is faster. This means that, all else being equal, when a market order does not change the price the persistence of the sign of subsequent orders is much stronger than when it does change the price.

We now study how the identity of the broker affects this behavior.
We look at the probability that the signs of the orders at $t$ and $t+\tau$ are the same, conditioned on whether they were placed by the same broker and also on whether the event at $t$ changed the price or not.
We study the following probabilities:
\begin{eqnarray}
P(\epsilon^i_t=\epsilon^j_{t+\tau}~|~i=j~;~MO_t^0)~~~~~~~~~P(\epsilon^i_t=\epsilon^j_{t+\tau}~|~i\neq j~;~MO_t^0)\nonumber\\
P(\epsilon^i_t=\epsilon^j_{t+\tau}~|~i=j~;~MO_t')~~~~~~~~~P(\epsilon^i_t=\epsilon^j_{t+\tau}~|~i\neq j~;~MO_t')
\label{probabilities}
\end{eqnarray}

\begin{figure}[ptb]
\begin{center}
\includegraphics[width=0.49\textwidth,angle=0]{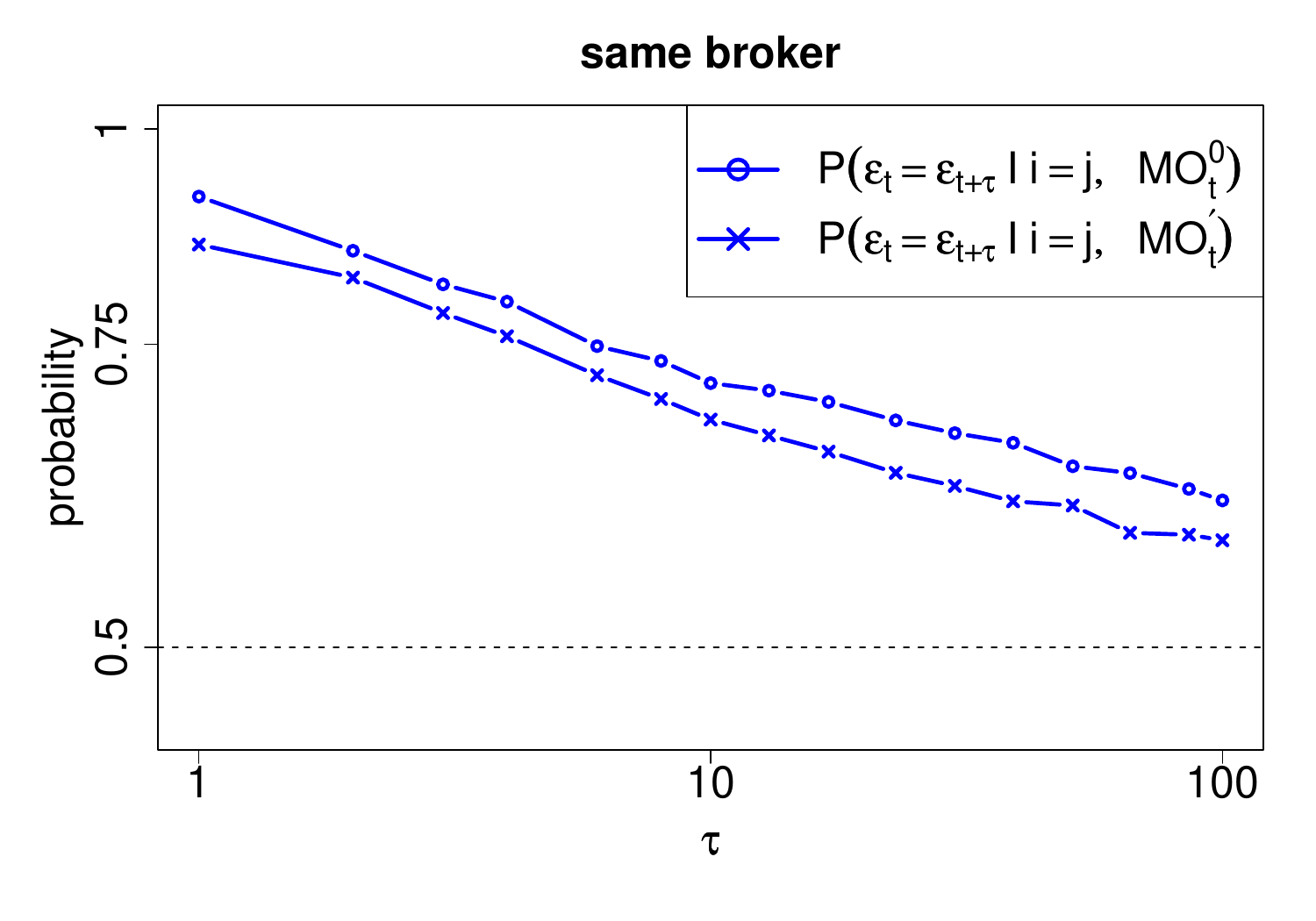}
\includegraphics[width=0.49\textwidth,angle=0]{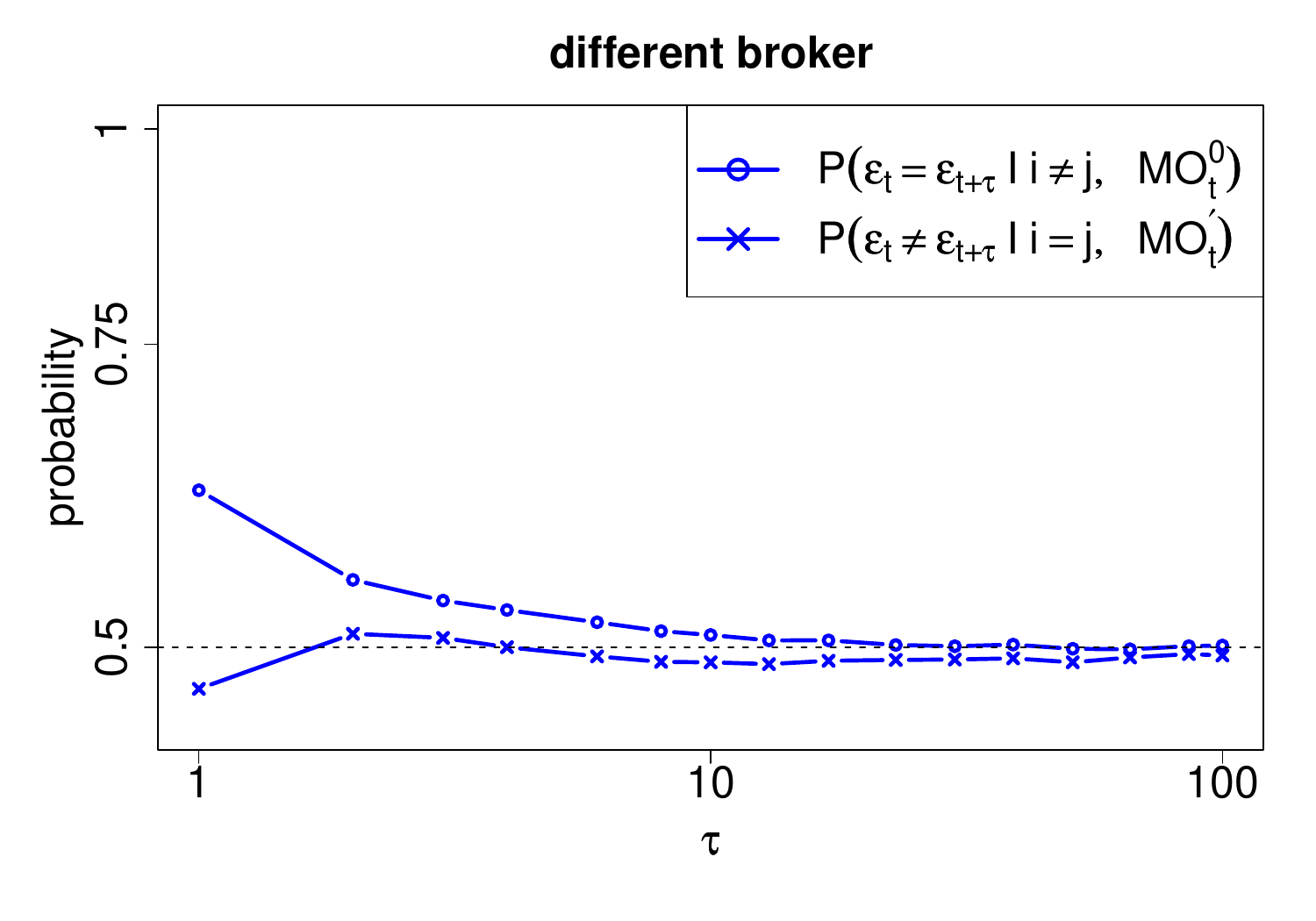}
\end{center}
\caption{\label{fab} The probability that market order signs separated by lag $\tau$ are the same, conditioned on whether the members placing the orders are the same and whether or not the first market order changes the price, as described in Eq.~(\ref{probabilities}).  \textit{Left panel}: $i=j$: The same member places the order at $t$ and $t+\tau$.
If the market order at time $t$ changes the price ($MO_t^{'}$, represented by "x") the behavior is nearly the same as if it does not change the price  ($MO_t^0$, represented by "o"), other than a slight shift in scale.
\textit{Right panel}: $i\neq j$: Different members place the order at $t$ and $t+\tau$. 
Whether or not the market order at time $t$  changes the price now makes a big difference in the behavior.   After an event that does not change the price, other brokers tend to put their order in the same direction with a probability higher than $0.5$.  After an event that changes the price, in contrast, at most lags it is slightly more probable that other brokers put their orders in the opposite direction. Data is for AZN for 2000-2009.}
\end{figure}

In Figure \ref{fab} we plot the probabilities above on a double logarithmic scale.  When the broker at time $t$ and $t+\tau$ is the same there is no qualitative difference, regardless of whether the order at time $t$ changed the price, other than a small shift in scale.

In contrast, when different brokers place the orders at time $t$ and $t+\tau$, the behavior changes dramatically.   If the market order at time $t$ does not change the price, the probability that the signs of the two orders are the same is lower than before, but still higher than $1/2$.  In contrast, if the market order at time $t$ changes the price, for most lags $\tau$ the subsequent order is more likely to have the opposite sign.
Thus, after a market order that does not change the price, other brokers tend to place their orders in the same direction, indicating a slight herding on short time scales.  In contrast, after a market order that changes the price, the opposite happens: they are more likely to put their orders in the opposite direction, causing negative autocorrelations.

\begin{figure}[ptb]
\begin{center}
\includegraphics[width=0.49\textwidth,angle=0]{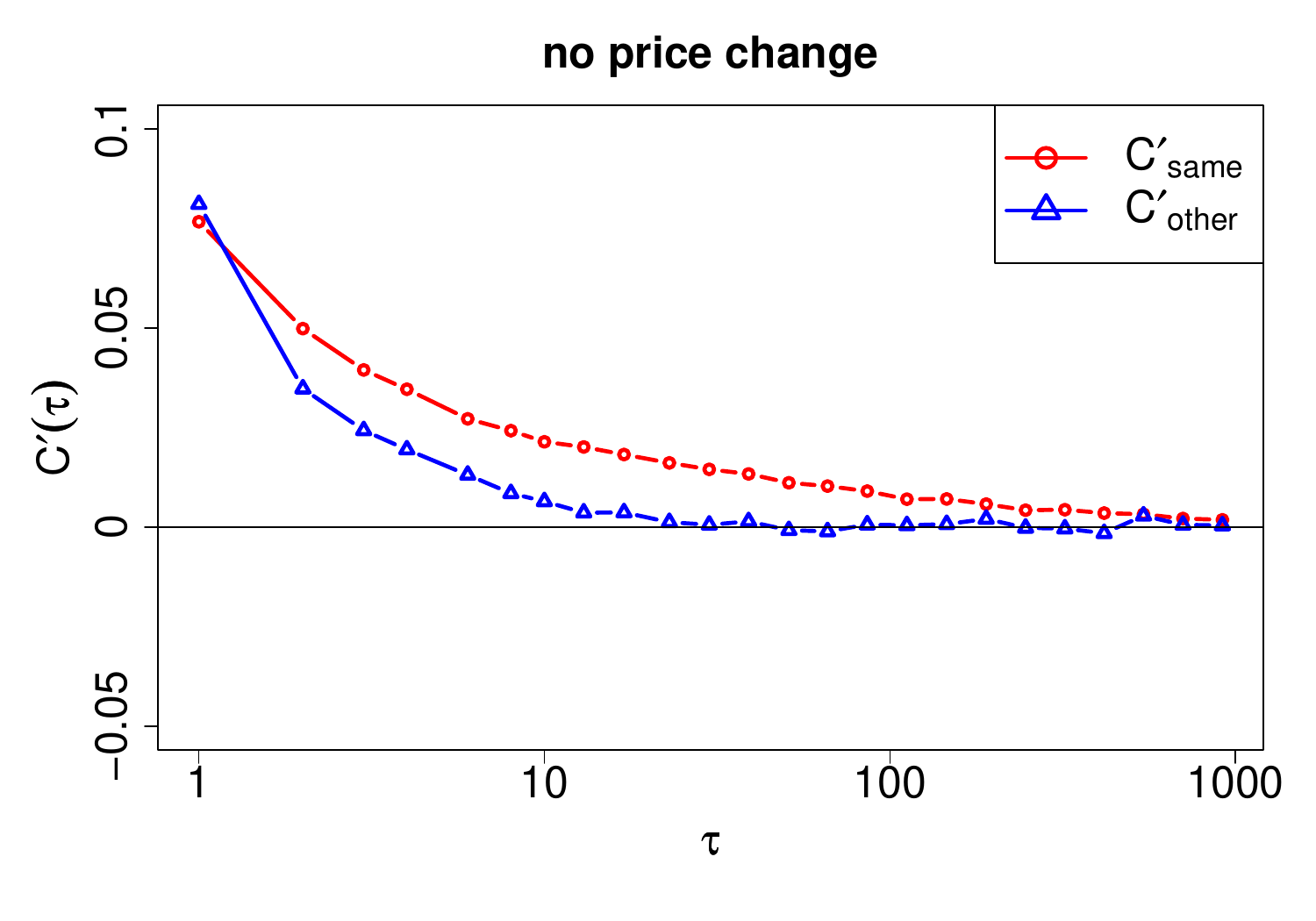}
\includegraphics[width=0.49\textwidth,angle=0]{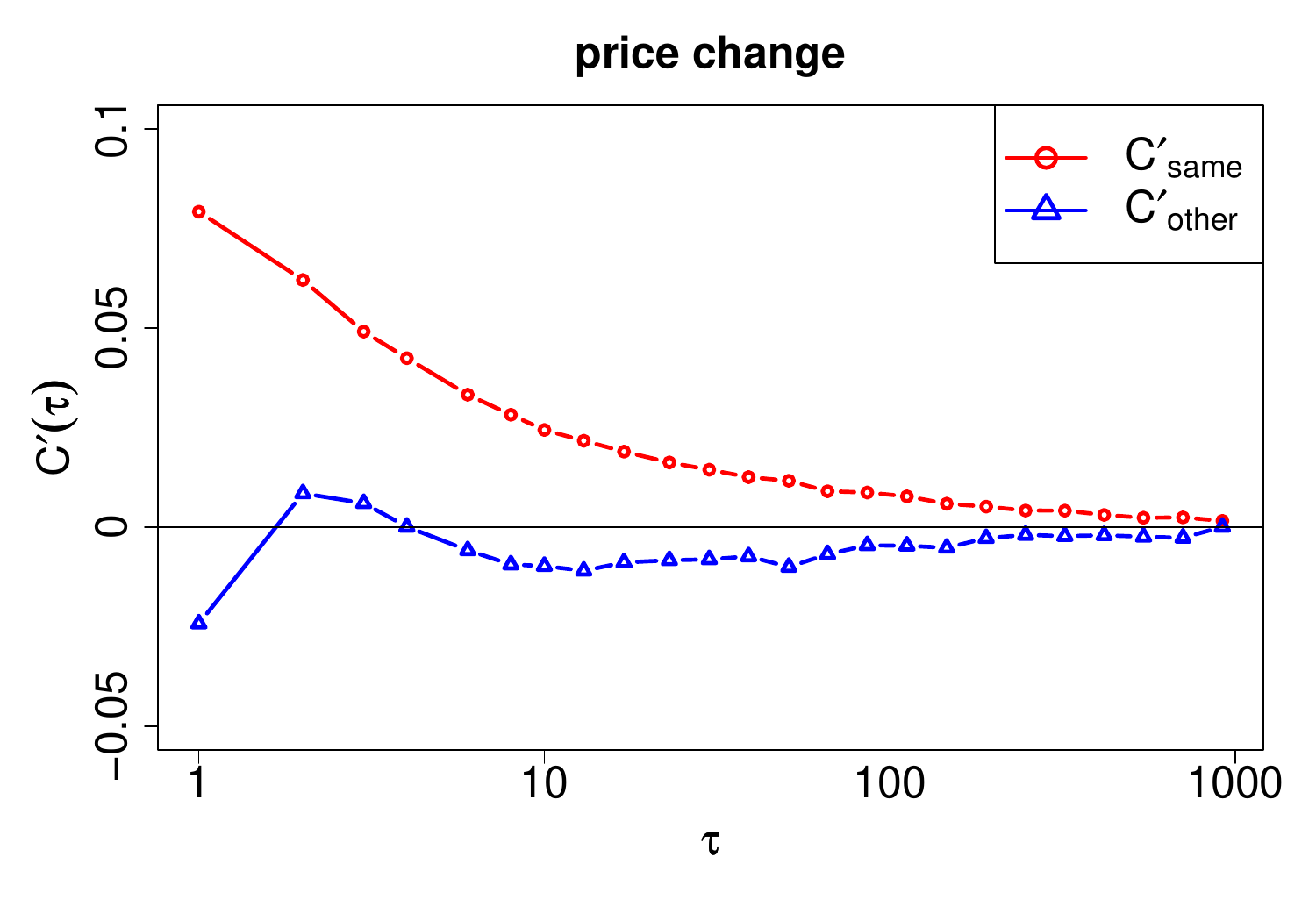}
\end{center}
\caption{\label{corrcond}  $C_{same}(\tau)$ vs. $C_{other}(\tau)$ components of the autocorrelation function of market order signs conditional on the event at time $t$.  The left panel is conditioned on the case where the original market order does not change the price, and the right panel is the case where it does.  The negative contribution to $C_{other}(\tau)$ is entirely due to the latter case. Data is for AZN for 2000-2009.}
\end{figure}

This is confirmed by decomposing the autocorrelation conditional on whether or not the market order at time $t$ changed the price, shown in Figure \ref{corrcond}.
More precisely, we apply the decomposition of Eq.~(\ref{components}) to the conditional autocorrelations $E[(\epsilon_t -\mu)(\epsilon_{t+\tau}-\mu)|MO_t^0]$ and $E[(\epsilon_t -\mu)(\epsilon_{t+\tau}-\mu)|MO_t']$, where $\mu$ is the mean market order sign.  When the market order at $t$ does not change the price, $C_{other}(\tau)$ decays to zero but never becomes very negative; in contrast, when the market order at $t$ changes price, it is negative for roughly $\tau > 5$. The shape of $C_{other}(\tau)$ is qualitatively similar to the probabilities shown in Figure \ref{fab}.

The conclusion is that the observed negative contribution of $C_{other}(\tau)$ is related to the difference in the response of agents to market orders placed by others, depending on whether or not they changed the price.  When a market order does not cause a price change, agents continue being more likely to place orders of the same sign, regardless of who placed the original market order.  In contrast, if a market order triggers a price change, other agents place fewer market orders in the same direction than in the opposite direction.  The interesting point is that splitting traders seem to act like ``noise traders'', i.e. they do not adapt their behavior to their own impact. The reason for this could be that they have already calculated their impact in their estimates of the trading cost, so it is expected. For a discussion on the possible microstructural mechanims see also T\'oth et al. (\citeyear{Toth11}) and Taranto et al. (\citeyear{Taranto14}). We are still far from a fully satisfactory explanation of the large variety of limit order book mechanisms responsible for the observed behavior. One possible explanation is that when the price moves up, for example, other agents, which were previously buying with market orders, switch to buying with limit orders placed into the enlarged spread. This increases the bid, making the use of market orders more favorable for sellers. The combined effect of the decrease of buyers using market orders and the increase of sellers facing a more favorable price could explain the negative correlation observed in the right panel of Figure. \ref{corrcond}. 
\section{Conclusions}

We have shown that in the LSE the cause of the extreme persistence of order flow on short timescales (corresponding to 500 transactions or less, typically about an hour) is overwhelmingly due to autocorrelated trading by individual members of the exchange rather than interactions between them. The fact that bursts of orders of the same sign mainly come from a single member, rather than several members, suggests that on these timescales herding is dominated by order splitting. We observe behavior at very short time scales that can be interpreted as herding, but for longer times this goes to zero, and indeed, we often observe that the order flow of a given exchange member is actually anti-correlated with that of others.  We should make the caveat that we are only able to obtain statistically significant results for fairly short time scales of less than a few hours. This makes it quite possible that herding is a stronger factor on longer timescales.

Our analysis is based on a decomposition of the autocorrelation function into a component due to cross-correlation of different investors and a component due to autocorrelation of investors with themselves.  We apply this to data identifying the members of the exchange, who often act as brokers for other investors.  In order to understand whether these results apply to single investors we developed a set of models for investor behavior and brokerage choice.  We combine these into scenarios with different combinations of investor behavior and brokerage, and use them as null hypotheses.   The herding null hypotheses are strongly rejected by the data; in contrast, splitting is not rejected. We believe that this shows that, even though our analysis was done at the level of brokers, the main conclusion applies at the level of investors as well.

The methods that we have developed here have implications beyond this study.  Data with brokerage information is more widespread than data about investors.  The methods that we have developed here provide a proof of principle for how brokerage data can be used to infer properties of investor behavior.  For example, we show that under the extreme assumption of dynamically random choice of brokers, in which investors randomly choose new brokers at each time step, the autocorrelation of order flow gives the appearance of strong herding.  The fact that the brokerage data strongly favors order splitting over herding shows that real brokerage choices must be very different than the dynamically random model. This result makes it clear that investors do not randomize their brokerage choice, but are rather consistent in their choice of brokers. Instead the data is consistent with the idea that investors use only a small number of brokers who they are consistent with through time. This is in line with evidence from Linnainmaa and Saar (\citeyear{Linnainmaa12}) and Fong et al. (\citeyear{Fong14}).

Our results here are consistent with the hypothesis that the origin of long memory in order flow is mainly due to investors who consistently execute their large orders through at most a few brokers, splitting them into small pieces.  This is exactly what one expects from algorithmic execution engines, who take large orders, split them into pieces and execute them throughout the day through their own brokers.   Interestingly, however, despite the increase in usage of algorithmic brokers, we do not observe any increase in $C_{same}(\tau)$ vs. $C_{other}(\tau)$ during the ten year period of our study.
The behavior is extremely consistent across different stocks and time periods.  It is also fairly consistent across member firms.  While a few member firms have less directional persistence than others, the vast majority are quite persistent, and in an almost identical way.  Furthermore the persistence exhibits itself similarly in both trading direction and trading activity.

As discussed in the introduction, an important consequence of autocorrelated order flow is its effect on market impact.  As shown by Farmer et al. (\citeyear{Farmer11}) this can result in rather precise (and hence sharply testable) predictions for the functional form and dynamic behavior of market impact.  For this the origin of the persistence of order flow matters:  If it were due to herding the implications would be substantially different.

Under the interpretation that the strong positive autocorrelation of investors is due to order splitting, the fact that investors split their orders so strongly implies that typically the market is in a certain sense out of equilibrium.  That is, if investors revealed their intentions when they made decisions, rather than concealing their intentions and revealing them only gradually, prices at any given moment might be substantially different than observed prices. Understanding the implications of this remarkably robust and widespread phenomenon deserves closer attention.

\section*{Acknowledgments}

We thank Jean-Philippe Bouchaud, Andrew Lo and Terry Odean for useful comments, and  National Science Foundation grant 0624351 for support.  Any opinions, findings, and conclusions or recommendations expressed are those of the authors and do not necessarily reflect the views of the National Science Foundation. FL acknowledges financial support from the grant SNS11LILLB ``Price formation, agent's heterogeneity, and market efficiency''.

\ifx\undefined\BySame
\newcommand{\BySame}{\leavevmode\rule[.5ex]{3em}{.5pt}\ }
\fi
\ifx\undefined\textmd
\newcommand{\textmd}[1]{{\sc #1}}
\fi
\ifx\undefined\emph
\newcommand{\emph}[1]{{\em #1\/}}
\fi

\section*{Appendix}

\subsection*{Derivation of autocorrelation decomposition}
The derivation is a trivial decomposition of the autocorrelation function into its pieces when the order flow can be labeled by the identity of the agent $i$ placing the order.
\begin{eqnarray*}
C(\tau) & = & \frac{1}{N}\sum_{t} \epsilon_t \epsilon_{t+\tau}-\left(\frac{1}{N}\sum_t \epsilon_t\right)^2\\
 & = & \frac{1}{N}\sum_t \sum_{i,j} \epsilon_t^i \epsilon_{t+\tau}^j-\left(\frac{1}{N}\sum_t\sum_i \epsilon_t^i\right)^2.
\end{eqnarray*}
Since by definition $P^i = N^i/N$ and $P^{ij} = N^{ij}/N$, by interchanging the order of sums this can be rewritten as
\begin{eqnarray*}
C(\tau) & = & \sum_{i,j} P^{ij}(\tau)\left[\frac{1}{N^{ij}(\tau)}\sum_t \epsilon_t^i \epsilon_{t+\tau}^j\right]-\left(\sum_iP^i\frac{1}{N^i}\sum_t\epsilon_t^i\right)^2\\
& = & \sum_{i,j}\left(P^{ij}(\tau)\left[\frac{1}{N^{ij}(\tau)}\sum_t \epsilon_t^i \epsilon_{t+\tau}^j\right]-P^iP^j\left[\frac{1}{N^i}\sum_t\epsilon_t^i\right]\left[\frac{1}{N^j}\sum_t\epsilon_t^j\right]\right).\\
\end{eqnarray*}
The autocorrelation function can then be written
\begin{equation}
\nonumber
C(\tau)=\sum_{i,j} P^{ij}(\tau)C^{ij}(\tau)+\sum_{i,j} \tilde P^{ij}(\tau) \mu^i \mu^j .
\end{equation}

\subsection*{Distortion of order flow decomposition by brokerage} 

Defining all notation in analogy with that for single investor order flow, for convenience assume $\mu'^i = 0$ for all $i$.  Then the autocorrelation for the order flow between brokerage $i$ at time $t$ and brokerage $j$ at time $t + \tau$ is
\begin{equation}
\label{brokerageDecomp1}
C'^{ij}(\tau) =  \frac{1}{N'^{ij}(\tau)} \sum_{t}^{}  \epsilon_t'^{i} \epsilon_{t + \tau}'^j.
\end{equation}
Using the definition of the brokerage map, Eq.~(\ref{brokerageMapDefinition}), and for convenience not explicitly stating the dependence of all the terms on $\tau$, this can be rewritten in terms of single investor order flow as
\begin{equation}
C'^{ij} =  \frac{1}{N'^{ij}} \sum_{kl} N^{kl} B_{ik} B_{jl}  \frac{1}{N^{kl}} \sum_t  \epsilon_t^{k} \epsilon_{t + \tau}^l = \frac{N}{N'^{ij}} \sum_{kl}  B_{ik} B_{jl} P^{kl} C^{kl}.
\end{equation}
The relation
\begin{equation}
C'(\tau)=\sum_{i,j} P'^{ij}(\tau)C'^{ij}(\tau),
\label{brokerageDecomp2}
\end{equation}
is analogous to Eq.~(\ref{ptilde}) for single investor order flow and is true by similar arguments.  By definition $P'^{ij} = N'^{ij}/N'$ and the total number of orders coming out of brokerages is the same as that for single investors, i.e. $N' = N$.  Substituting all of these into Eq.~(\ref{brokerageDecomp2}) gives
\begin{equation}
C'(\tau)=\sum_{i,j,k,l} B_{ik} B_{jl} P^{kl} C^{kl}.
\end{equation}


\end{document}